\documentclass[twocolumn]{aastex631}

\def\lapp{\ifmmode\stackrel{<}{_{\sim}}\else$\stackrel{<}{_{\sim}}$\fi}
\def\gapp{\ifmmode\stackrel{>}{_{\sim}}\else$\stackrel{>}{_{\sim}}$\fi}
\usepackage{graphicx}
\usepackage{multirow}
\usepackage{color}
\usepackage{amsmath}
\usepackage{soul}
\usepackage{hyperref}
\newcommand{\fluxcgs}{\ensuremath{\mathrm{erg}\,\mathrm{s}^{-1}\,\mathrm{cm}^{-2}}}

\shorttitle{}
\shortauthors{}
\begin{document}

\title{X-ray characterization of the pulsar PSR~J1849$-$0001 and its wind nebula G32.64+0.53 associated
with TeV sources detected by H.E.S.S., HAWC, Tibet AS$\gamma$, and LHAASO}

\correspondingauthor{Hongjun An}
\email{hjan@cbnu.ac.kr}
\author[0000-0003-0226-9524]{Chanho Kim}
\author[0000-0002-9103-506X]{Jaegeun Park}
\affiliation{Department of Astronomy and Space Science, Chungbuk National University, Cheongju, 28644, Republic of Korea}
\author[0009-0001-6471-1405]{Jooyun Woo}
\author{Sarah Silverman}
\affiliation{Columbia Astrophysics Laboratory, 550 West 120th Street, New York, NY 10027, USA}
\author[0000-0002-6389-9012]{Hongjun An}
\affiliation{Department of Astronomy and Space Science, Chungbuk National University, Cheongju, 28644, Republic of Korea}
\author[0000-0003-0890-4920]{Aya Bamba}
\affiliation{Department of Physics, Graduate School of Science, The University of Tokyo, 7-3-1 Hongo, Bunkyo-ku, Tokyo 113-0033, Japan}
\affiliation{Research Center for the Early Universe, School of Science, The University of Tokyo, 7-3-1 Hongo, Bunkyo-ku, Tokyo 113-0033, Japan}
\affiliation{Trans-Scale Quantum Science Institute, The University of Tokyo, Tokyo  113-0033, Japan}
\author[0000-0002-9709-5389]{Kaya Mori}
\affiliation{Columbia Astrophysics Laboratory, 550 West 120th Street, New York, NY 10027, USA}
\author[0000-0002-5365-5444]{Stephen P. Reynolds}
\affiliation{Physics Department, NC State University, Raleigh, NC 27695, USA}
\author[0000-0001-6189-7665]{Samar Safi-Harb}
\affiliation{Department of Physics and Astronomy, University of Manitoba, Winnipeg, MB R3T 2N2, Canada}
%\collaboration{(AAS Journals Data Scientists collaboration)}

\begin{abstract}
We report on the X-ray emission properties of the pulsar PSR~J1849$-$0001 and its wind nebula (PWN), as measured by Chandra, XMM-Newton, NICER, Swift, and NuSTAR. In the X-ray data, we detected the 38-ms pulsations of the pulsar up to $\sim$60\,keV with high significance. Additionally, we found that the pulsar’s on-pulse spectral energy distribution displays significant curvature, peaking at $\approx$60\,keV. Comparing the phase-averaged and on-pulse spectra of the pulsar, we found that the pulsar’s off-pulse emission exhibits a spectral shape that is very similar to its on-pulse emission. This characterization of the off-pulse emission enabled us to measure the $>$10\,keV spectrum of the faint and extended PWN using NuSTAR’s off-pulse data. We measured both the X-ray spectrum and the radial profiles of the PWN's brightness and photon index, and we combined these X-ray measurements with published TeV results. We then employed a multizone emission scenario to model the broadband data. The results of the modeling suggest that the magnetic field within the PWN is relatively low ($\approx 7\mu \rm G$) and that electrons are accelerated to energies $\gapp 400$\,TeV within this PWN. The electrons responsible for the TeV emission outside the X-ray PWN may propagate to $\sim$30\,pc from the pulsar in $\sim$10\,kyr.
\end{abstract}

\bigskip 
\section{Introduction}
\label{sec:intro}
It is well known that high-energy electrons exist in pulsar wind nebulae (PWNe), as evidenced by detections of very high-energy ($\ge$100\,GeV) photons emitted from them. Indeed, numerous H.E.S.S. and LHAASO sources detected at energies above 10\,TeV are associated with X-ray PWNe \citep[e.g.,][]{HESSHGPS2018,LHAASO2023}. Theories suggest that electrons and positrons are accelerated at the termination shock (TS), which is formed by the interaction between a pulsar's wind and the ambient medium \citep[][]{kc84b}. These energetic leptons propagate outward from the TS by advection and diffusion \citep[e.g.,][]{deJager2009}, generating a bubble of synchrotron radiation, i.e., a PWN. Electrons injected by old pulsars would have sufficient time to travel large distances from the pulsars, and these electrons are believed to contribute to the formation of TeV halos around aged pulsars \citep[][]{HAWC_halo2017,HESS2023}. Additionally, PWNe may play a role in generating the high-energy cosmic-ray electrons/positrons detected on Earth \citep[e.g.,][]{Pamela2009,Manconi2020}.

While young systems can accelerate electrons to very high energies \citep[][]{Cao2021}, due to radiative cooling, these electrons would not retain their energies when they escape from the compact X-ray emission region. Electrons that emit TeV photons via the inverse-Compton scattering process are expected to cool within a few kyrs
if the magnetic field ($B$) within the PWN is strong \citep[e.g., $B=20$--100\,$\mu$G in young PWNe;][]{Bamba2010a,Torres2014}. Consequently, electrons generated during the early stages of PWN evolution (when $B$ was higher) cannot account for the high-energy electrons responsible for creating TeV halos and the energetic cosmic-ray electrons detected on Earth. On the other hand, low-$B$ PWNe associated with middle-aged (characteristic age $\tau_c$ of 10--100\,kyr) pulsars with high spin-down power ($\dot E_{\rm SD}>10^{36}\rm \ erg\ s^{-1}$) are good candidates for producing high-energy electrons/positrons within the Galaxy. 

Recent studies have provided evidence of PeV electrons within several middle-aged PWNe \citep[e.g.,][]{Cao2021,Burgess2022,Woo2023} through the modeling of their broadband spectral energy distributions (SEDs). 
These high-energy electrons propagate outward, and they eventually escape from the PWNe and are injected into the ISM.
However, the injection rates and energetics of electrons remain unclear. These quantities depend on the flow properties and energy-loss mechanisms within PWNe \citep[e.g.,][]{Reynolds2016}, which can be investigated by modeling spatially varying properties and broadband SEDs of PWNe \citep[e.g.,][]{Park2023a} across different evolutionary stages. Of particular significance is investigating the most energetic electrons ($>100$ TeV), which is most effectively accomplished by analyzing their synchrotron X-ray emission, while the inverse-Compton emission in TeV energies is suppressed by the Klein-Nishina effect. Therefore, studying broad-band X-ray emission properties of PWNe with proper multi-zone modeling is essential, especially for those associated with energetic pulsars and TeV sources \citep[e.g.,][]{Mori2021}.

The bright X-ray source IGR~J1849$-$0000 was discovered by INTEGRAL
\citep[][]{Molkov2004}. A follow-up XMM-Newton observation \citep[][]{Terrier2008} resolved the INTEGRAL source to a point
source and extended emission ($\approx 150''$),
suggesting a pulsar+PWN system. 
The pulsar hypothesis was substantiated by the detection of a 38-ms pulsation 
\citep[PSR~J1849$-$0001; J1849 hereafter;][]{Gotthelf2011} and  $\dot P=1.42\times 10^{-14}\rm \ s\ s^{-1}$. The measured $P$ and $\dot P$ values imply a surface magnetic field $B_S=7.5\times10^{11}$\,G,
spin-down power $\dot E_{\rm SD}=9.8\times 10^{36}\rm \ erg\ s^{-1}$, and characteristic
age $\tau_c=43$\,kyr, thus categorizing the point source as an energetic and middle-aged pulsar. 
Furthermore, this middle-aged pulsar and its PWN are of particular interest as
their emission is significantly detected above 100\,TeV by
HAWC \citep[][]{3HWC2020}, LHAASO \citep[][]{LHAASO2023}, and Tibet air shower array \citep[Tibet AS$\gamma$;][]{Amenomori2023}
with the maximum photon energy of 350\,TeV \citep[][]{LHAASO2021}.
This indicates the presence of energetic particles within the PWN
(G32.64+0.53\footnote{http://snrcat.physics.umanitoba.ca/}; G32.6 hereafter).

\citet{kh15} conducted an in-depth analysis of Chandra, XMM-Newton, RXTE, and INTEGRAL data of the J1849+G32.6 system and found that the pulsed spectrum of the pulsar is well described by a power-law model with a hard photon index of $\Gamma = 1.37$. They further compared this `pulsed (on$-$off)' spectrum with the total (on+off) spectrum measured with INTEGRAL, as \citet{Terrier2008} did. Based on this comparison, both of these previous studies suggested that the pulsar exhibits a curved spectrum.
Additionally, \citet{kh15} observed that the PWN emission in the inner region, as measured with the Chandra data, exhibits a harder spectrum compared to the spectrum from the outer region, as measured with the XMM-Newton data. They attributed this spectral change to the synchrotron burn-off effect. 

 Further exploration of the spatially varying PWN emission can give insights into particle transport mechanisms within the source \citep[e.g.,][]{Tang2012,Porth2016,KimS2020}, subsequently exploring how energetic electrons are injected into the ISM. Moreover, the spectral softening could potentially manifest as a high-energy spectral break in the spatially-integrated spectrum; NuSTAR's hard X-ray data may facilitate detecting a spectral break \citep[e.g.,][]{Nynka2014,Madsen2015} or a cut-off \citep[e.g.,][]{An2019} at energies $\ge$10\,keV. Furthermore, a precise characterization of the putative curvature in J1849's spectrum can offer insights into the mechanism of pair production in rotation-powered pulsars \citep[][]{Harding2017}.

In this paper, we carry out an X-ray characterization of the emissions from the pulsar 
J1849 and its wind nebula G32.6 using archival X-ray observations as well as new NuSTAR data.
We present the results of our X-ray data analyses
in Section~\ref{sec:sec2} and construct a broadband SED
of the PWN by supplementing our X-ray measurements with the TeV results of H.E.S.S., HAWC,
LHAASO, and Tibet AS$\gamma$ (Section~\ref{sec:sec3}). We employ a multi-zone emission model
to interpret the broadband data and infer the properties of this PWN (Section~\ref{sec:sec3}).
We discuss the results in Section~\ref{sec:sec4} and present a summary in Section~\ref{sec:sec5}.
The uncertainties reported in this paper correspond to 1$\sigma$ confidence intervals unless otherwise noted.
 
\section{X-ray Data Analysis}
\label{sec:sec2}

\subsection{Data reduction}
\label{sec:sec2_1}
We analyzed the X-ray observation data of J1849 obtained by Chandra, XMM-Newton, NICER, Swift, and NuSTAR.
We reprocessed the Chandra ACIS-S data acquired on 2012 Nov. 16 for 23\,ks (Obs. ID 13291)
using the {\tt chandra\_repro} tool of CIAO~4.14.
The XMM-Newton observation (2011 Mar. 23; Obs. ID 0651930201) was carried out
with the full-frame and small-window mode for the MOS and the PN detectors,
respectively. We processed these data with the {\tt emproc} and {\tt epproc} tools
of SAS~20211130\_0941, and we further removed particle flares following the
standard procedure. Net XMM-Newton exposures after the data screening are 47\,ks, 50\,ks, and 37\,ks for MOS1, MOS2, and PN, respectively.
The NICER data were collected between 2018 Feb. 13 and 2022 Jul. 14
(143 observations with the Obs. IDs 1020660101--5505050501).
We processed the NICER observations using the {\tt nicerl2} script
integrated in HEASOFT~v6.31.
We also used the Swift windowed-timing mode data taken on 2017 Mar. 19 for 12.5\,ks
(Obs. ID 00034978002) to extend the baseline for our pulsar timing study
(Section~\ref{sec:sec2_2}). The Swift data were processed
with the {\tt xrtpipeline} tool.
The source was observed by NuSTAR on 2020 Nov. 24, and we processed the data using {\tt nupipeline}
along with the {\tt strict} SAA filter. The net
exposure of the NuSTAR observation is 51\,ks for each of FPMA and FPMB.

\subsection{Pulsar timing analysis}
\label{sec:sec2_2}
\begin{figure}
\centering
\includegraphics[width=3.05 in]{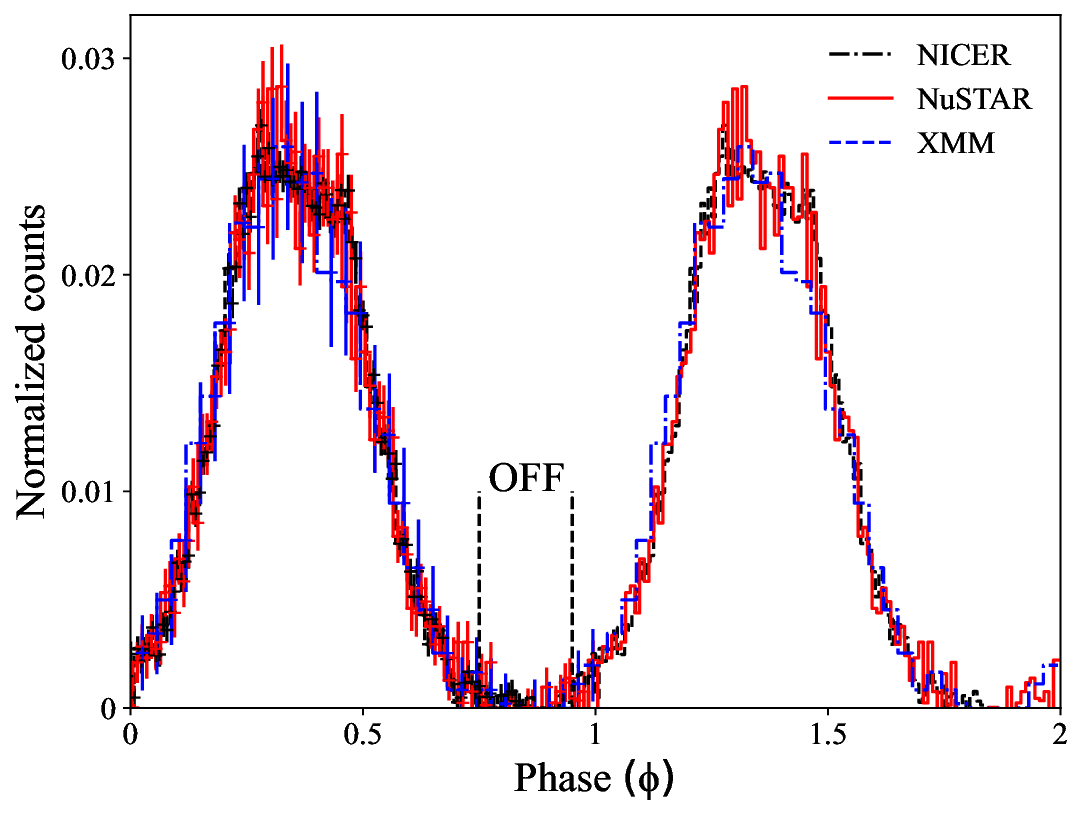} \\
\includegraphics[width=3.05 in]{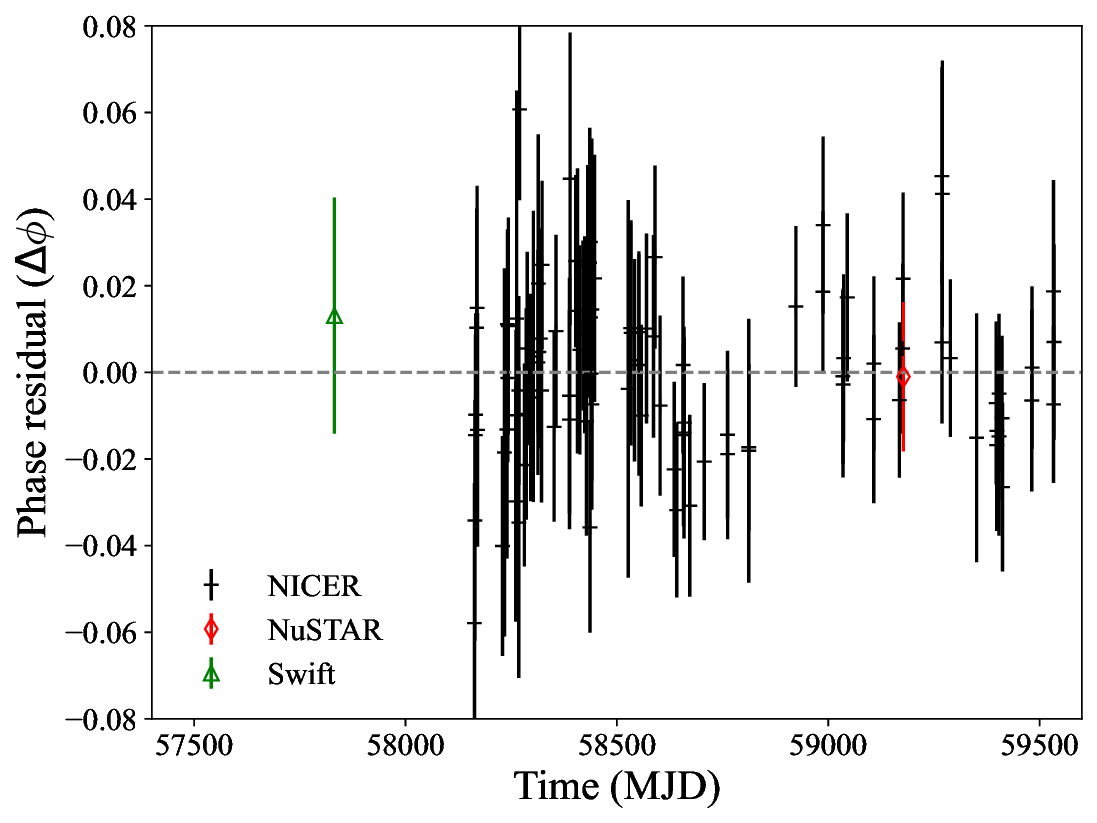} \\
\figcaption{{\it Top}:
Pulse profiles in the 1--10\,keV (blue; XMM-Newton), 2--8\,keV (black; NICER), and 3--60\,keV
(red; NuSTAR FPMA and FPMB combined) bands.
Backgrounds and off-pulse emissions are subtracted from the profiles, and each profile is
normalized to have an integrated count of 1. The vertical dashed lines denote the off-pulse interval ($\phi$=0.75--0.95).
The reference phase ($\phi=0$) of the XMM-Newton profile was adjusted to align
with the NICER and NuSTAR profiles. Two cycles are shown for clarity.
{\it Bottom}: Phase residuals after optimizing the timing solution (see text).
\label{fig:fig1}}
\vspace{0mm}
\end{figure}

\begin{table}[t]
\vspace{-0.0in}
\begin{center}
\caption{Pulsar timing parameters}
\label{ta:ta1}
\vspace{-0.05in}
\scriptsize{
\begin{tabular}{lc} \hline\hline
Parameter				& Value   \\ \hline
Range of dates (MJD)			& 57831--59774   \\
Epoch (MJD TBD)				& 58239.91628621966   \\
Frequency (Hz)				& 25.9590178608(4)   \\
1st derivative (Hz $\rm s^{-1}$)	& $-9.54076(2)\times 10^{-12}$   \\
2nd derivative (Hz $\rm s^{-2}$)	& $1.74(2)\times 10^{-22}$   \\
3rd derivative (Hz $\rm s^{-3}$)	& $-3.16(9)\times 10^{-30}$   \\
4th derivative (Hz $\rm s^{-4}$)	& $3.9(2)\times 10^{-38}$   \\ \hline
\end{tabular}}
\end{center}
\vspace{-0.5 mm}
\end{table}

A timing solution for J1849, valid between MJDs 57830.9 and 58391.0, was reported by \citet{Bogdanov2019} \citep[see also][]{Ho2022}. We extend this solution to cover the 5.4-year period spanning the epochs of the Swift, NICER, and NuSTAR observations. In the Swift data, we extracted events within an $R=16''$ circle centered at the pulsar in the 1--6\,keV band, while for the NuSTAR data, we used an $R=60''$ extraction circle in the 3--60\,keV band. Because the source pulsations were not well detected by NICER below 2\,keV, we employed the 2--8\,keV band for the NICER analysis. Our analysis included 108 NICER observations (for both timing and spectral analyses) in which the pulse profiles were confidently measured. We applied barycenter corrections to the photon arrival times, utilizing the source position (R.A., decl.)=($282.2568023^\circ$, $-0.0216153^\circ$), and carried out a semi-phase-coherent timing analysis \citep[e.g.,][]{AnArchibald2019}.

\begin{figure*}
\centering
\hspace{-3.0 mm}
\begin{tabular}{ccc}
\includegraphics[width=2.2 in]{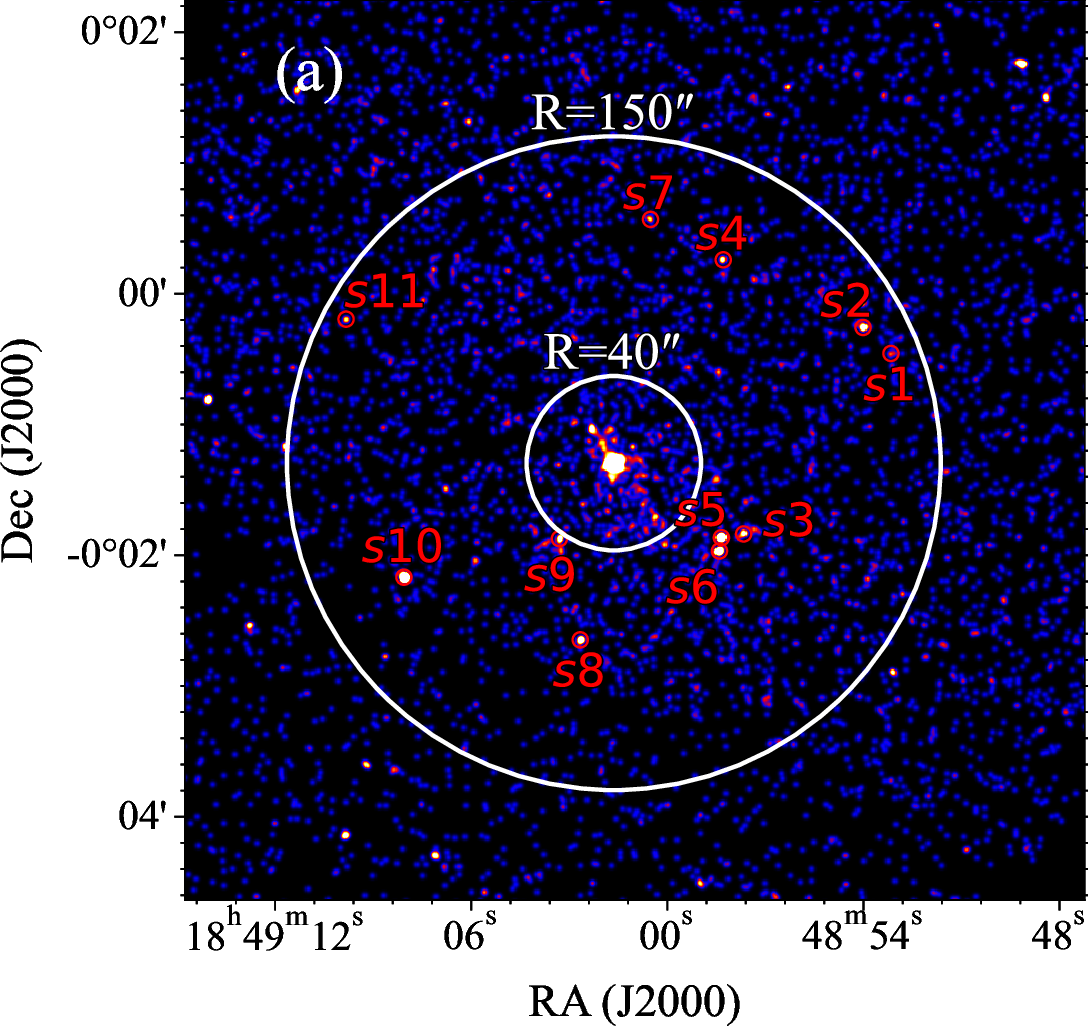} &
\includegraphics[width=2.2 in]{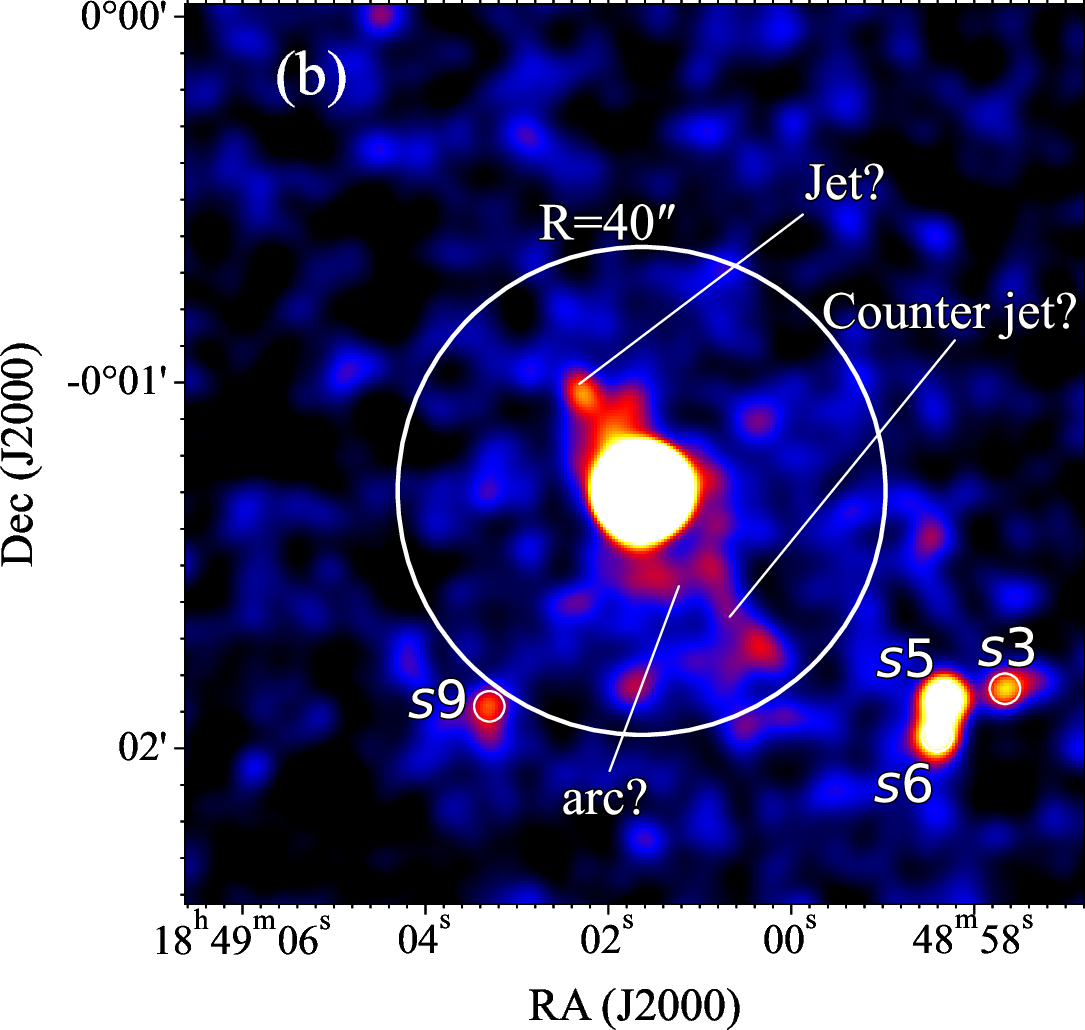} &
\includegraphics[width=2.2 in]{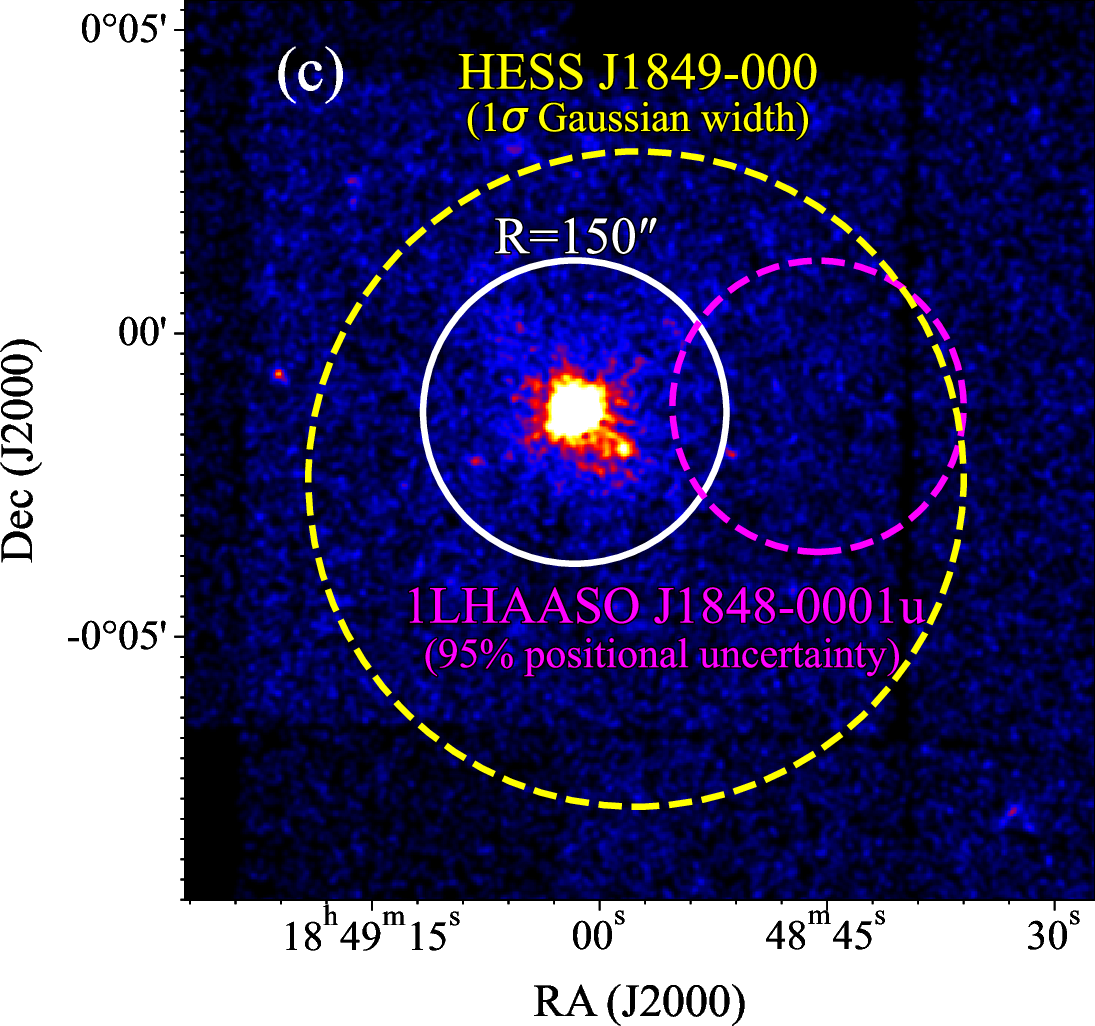}
\end{tabular}
\figcaption{1--7\,keV Chandra ACIS-S (panels (a) and (b)) and XMM-Newton MOS (panel (c)) count images of the pulsar J1849 and its PWN.
For legibility, the images are logarithmically scaled and smoothed with a Gaussian kernel
having 1$\sigma$ widths of 2.5, 4, and 1.3 pixels in panels (a), (b), and (c), respectively.
({\it a}) Red circles denote contaminating point sources within an $R=150''$ region. A 40$''$ circle is shown for reference. ({\it b}) In this zoomed-in Chandra image, a jet and counter-jet structure in the northeast-southwest direction and an arc structure are apparent, but the detection significance for these structures was low ($\le2.5\sigma$).
({\it c}) Faint and extended emission from the PWN is visible out to $R\lapp 150''$. A bright spot in the southwest of the pulsar is produced by the point sources S3, S5, and S6 (see panel (a)).
The H.E.S.S. source ($0.09^\circ$; 1$\sigma$ Gaussian width)
and the LHAASO ($0.04^\circ$; 95\% positional uncertainty) source are
denoted by yellow and magenta circles, respectively.
\label{fig:fig2}}
\vspace{0mm}
\end{figure*}

We constructed a pulse-profile template using the Swift and NICER data for which 
the existing solution is valid (MJD~57830.9--58391.0). We then fit the template with
a double-Gaussian function and used the function in our analysis.
We progressively folded the later observations and measured a phase shift of
each observation by fitting the observed pulse profile with the function.
In the case that notable drifts occurred in the pulse-arrival phases,
we updated the timing solution by introducing a higher frequency derivative
to ensure phase alignment. Following this, we created a new pulse-profile
template using the revised timing solution and updated the Gaussian function.
We repeated this procedure for all the Swift, NICER, and NuSTAR observations.
To ensure phase coherence across the Swift, NuSTAR, and 4.5\,yr NICER data,
it was necessary to incorporate four frequency derivatives.
The results of this semi-phase-coherent analysis are presented in Table~\ref{ta:ta1} and
Figure~\ref{fig:fig1}. The source pulsations were significantly detected
up to the 50--60\,keV band by NuSTAR with a chance probability of $6\times 10^{-4}$.

Because our timing solution is not valid at the much earlier XMM-Newton observation epoch due to the uncertainties in the timing parameters,
we performed $H$ tests \citep[][]{drs89} to measure the XMM-Newton pulse profile.
For this analysis, we used the 1--10\,keV band and an $R=16''$ circular region to extract source
events from the PN data. We conducted a search for pulsations around the previously reported period, while holding $\dot P$ fixed at the measured value of $1.4224\times 10^{-14}$
\citep[][]{Gotthelf2011,kh15}. The pulsations were detected with high significance with 
an $H$ statistic of $\sim$4000. The XMM-Newton-measured profile is displayed in Figure~\ref{fig:fig1}.
The on-pulse profiles  (off-pulse emission was subtracted using the phase interval $\phi$=0.75--0.95) measured with the three observatories agree with each other very well, and we further verified that the on+off profiles measured with XMM-Newton and NuSTAR, obtained by subtracting the background taken from circular regions ($R=32''$ for XMM-Newton and $R=60''$ for NuSTAR) at $R\sim150''$ from the pulsar, 
also agreed well with each other.
It is important to note that the pulsations were significantly detected in the $<$2\,keV band
of the PN data. This means that the non-detection of $<$2\,keV pulsations in the NICER data likely results from elevated background levels (e.g., flares) present in the data, rather than originating from intrinsic
emission properties of J1849 or the influence of strong Galactic absorption.

\subsection{Image analysis}
\label{sec:sec2_3}
While the extended PWN G32.6 has been previously identified
\citep[][]{Terrier2008, Gotthelf2011, kh15, Vleeschower2018}, a careful
image analysis of the Chandra, XMM-Newton, and NuSTAR data is required to assess contamination by the pulsar in the PWN emission region.
This is crucial for analyzing the low imaging resolution data collected
by XMM-Newton and NuSTAR. The optics vignetting effect
is not a concern for our investigation of the central region (e.g., $<30''$).
At $R=150''$, this effect was estimated to be $\sim$3--8\% for XMM-Newton and Chandra,
and $\sim$14\% for NuSTAR.

\begin{figure*}
\centering
\hspace{-3.0 mm}
\begin{tabular}{ccc}
\hspace{-3.0 mm}
\includegraphics[width=2.3 in]{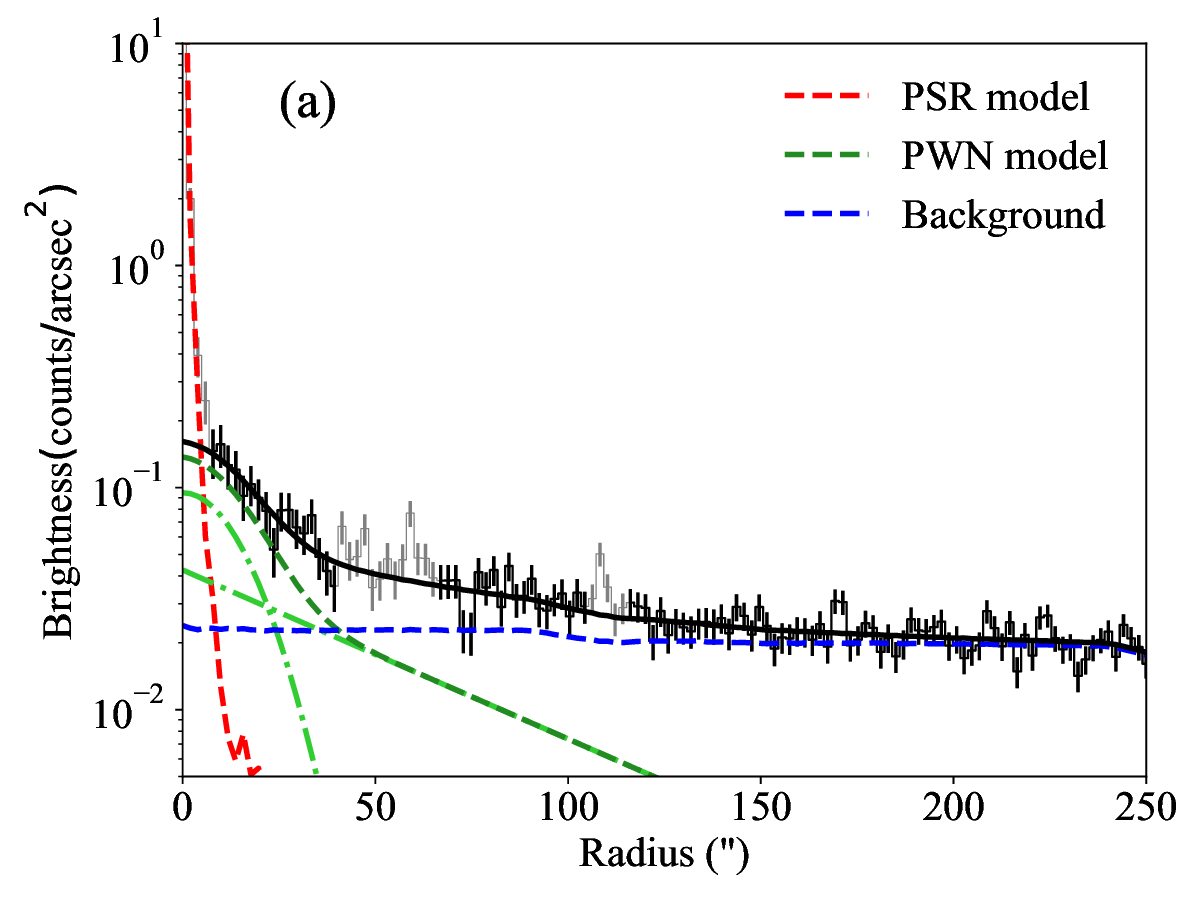} &
\hspace{-3mm}
\includegraphics[width=2.3 in]{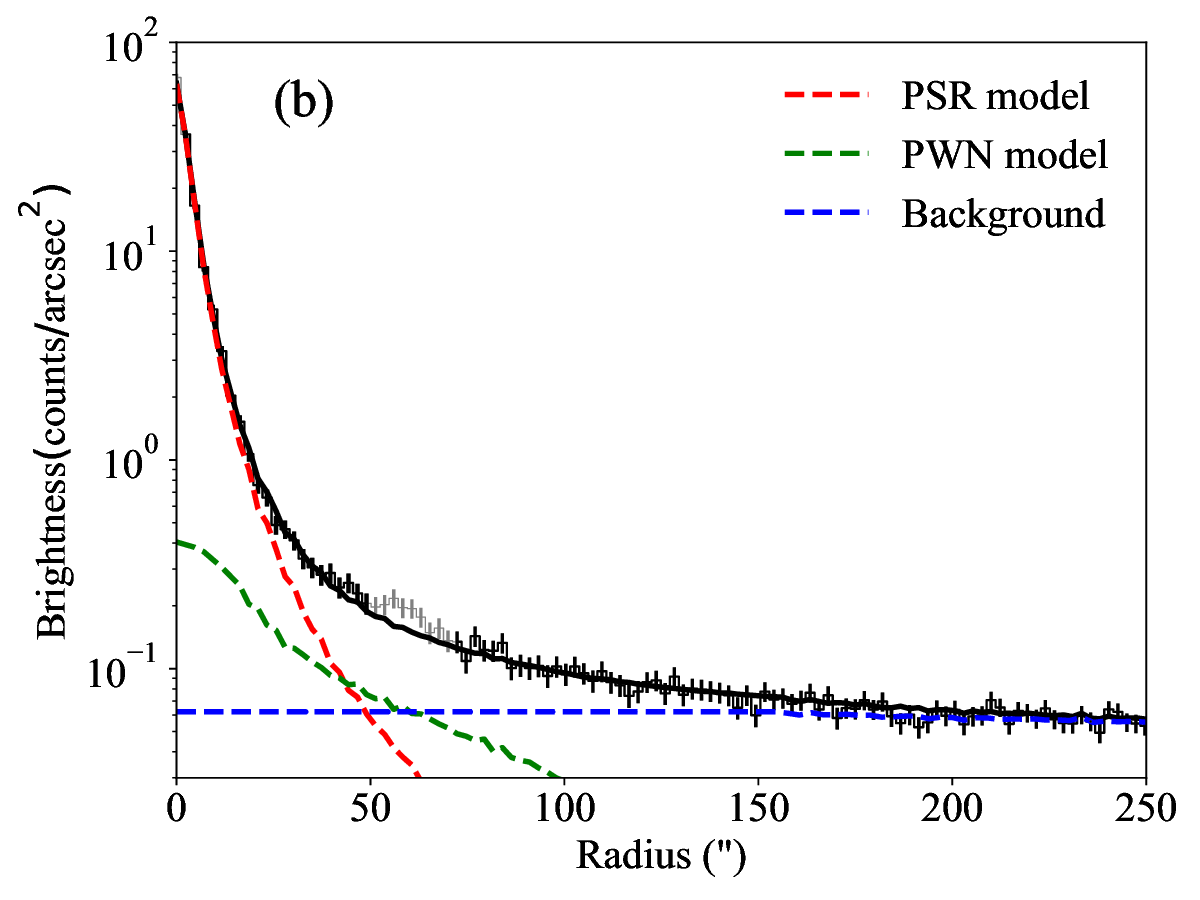} &
\hspace{-3mm}
\includegraphics[width=2.3 in]{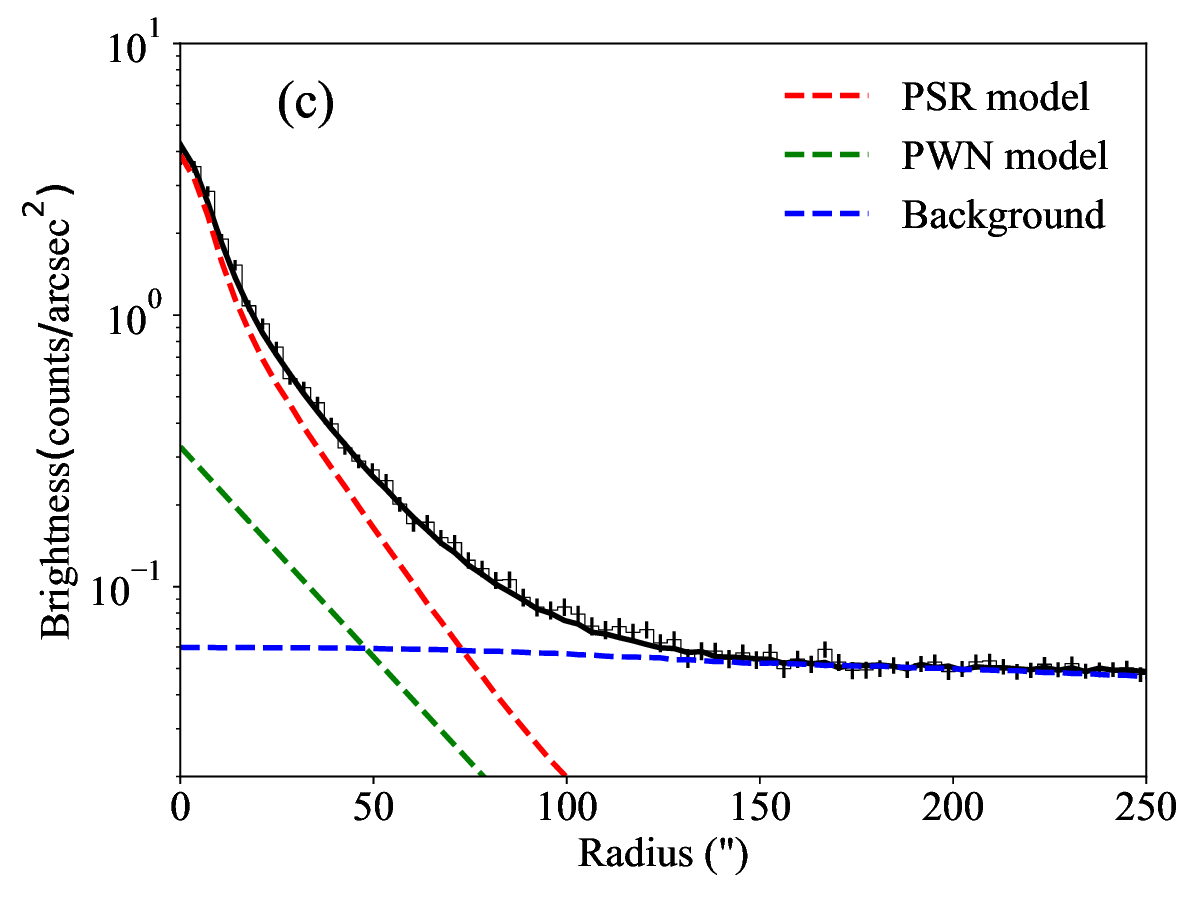} \\
\end{tabular}
\figcaption{Radial profiles measured with Chandra and XMM-Newton/MOS in the 2--7\,keV (panels (a) and (b)), as well as
with NuSTAR in the 5--50\,keV band (panel (c)).
The red, green, and blue curves correspond to the pulsar, PWN, and background models, respectively, and the solid black curve is the summed model.
In panel (a), the dot-dashed lines are the Gaussian and exponential functions used for the PWN model (see text).
The gray data points in panels (a) and (b) are ignored in our fits as they are contaminated by the piled-up pulsar ((a)) or other point sources ((a) and (b)).
\label{fig:fig3}}
\vspace{0mm}
\end{figure*}

\subsubsection{Analysis of the Chandra and XMM-Newton images}
\label{sec:sec2_3_1}
We present the 2--7\,keV count image measured by Chandra in panels (a) and (b) of Figure~\ref{fig:fig2}.
The central $R\lapp1''$ region is significantly affected by the pile-up effect caused by the bright pulsar.
The $R\lapp 40''$ region surrounding the pulsar appears brighter than other regions and displays some possible substructures. These include a jet-like feature in the northeast direction, a counterjet in the southwest direction (distinct from the readout trail), and an arc to the south of the pulsar.
However, each of these structures is statistically insignificant ($\le2.5\sigma$).

On a larger scale, the extended PWN with a radius of $R\approx 150''$ (Figure~\ref{fig:fig2} (c))
is detected by XMM-Newton, as previously reported \citep[e.g.,][]{Terrier2008}.
This X-ray PWN overlaps with the regions of TeV emission measured by
H.E.S.S and LHAASO \citep[][]{HESSHGPS2018,LHAASO2023}. HAWC has also identified a source with a $0.16^\circ$ offset in the north-east direction \citep[][]{3HWC2020}.

We further characterized the PWN emission using radial profiles around the pulsar position.
The Chandra profile (Figure~\ref{fig:fig3} (a)) exhibits J1849,
a narrow core, and a broad wing. The latter two components extend to large distances, 
representing the extended PWN.
The profile also shows spiky features that arise from contamination by point sources within the field.
To model the Chandra profile, we considered a uniform background (exposure map) and a PWN model, while excluding the piled-up pulsar ($R\le 6''$) and other contaminating sources. However, we found that a single-component model for the PWN emission failed to adequately fit the data. Consequently, we adopted a two-component function, a Gaussian combined with an exponential tail: $A \mathrm{exp}(-r^2/2\sigma^2) + B\mathrm{exp}(-r/l)$.
This function successfully fit the data, as illustrated in Figure~\ref{fig:fig3} (a), with
the best-fit parameters of $\sigma=10.8\pm2.2''$ and $l=49.5\pm 5.8''$.

We also carried out an analysis of the 1D radial profile derived from the XMM-Newton/MOS data (Figure~\ref{fig:fig3} (b)).
For the analysis of the profile, we utilized the MOS point spread functions (PSFs) generated using the
{\tt psfgen} tool of SAS to represent the pulsar emission.
To construct a PWN model, we convolved the Chandra PWN model with the XMM-Newton PSFs.
In this analysis, we adopted a flat background, i.e., the exposure map, 
and fit the profile by optimizing the normalization factors of the
PSF, PWN, and background components. In the fitting procedure, we excluded
regions that were contaminated by point sources, such as the small bump at $R\approx60''$.
The results are shown in Figure~\ref{fig:fig3} (b).
Notably, within a radius of $R\le50''$, the pulsar appears brighter than the PWN, and in the central $16''$ region, the estimated 2--10\,keV PWN counts amount to $<$2\% of the pulsar's count.

The XMM-Newton/PN small-window image, despite its limited coverage of the PWN,
allowed the distinction between the on- and off-pulse emission from the pulsar. We generated
radial profiles of the PN data utilizing the on- and off-pulse intervals defined in Figure~\ref{fig:fig1},
and then fit these profiles with the same model applied to the MOS data.
Our analysis revealed that the 2--7\,keV off-pulse emission from the pulsar (i.e., PSF)
is $22\pm1$\% of its on$-$off emission (per spin cycle).

\subsubsection{Analysis of the NuSTAR image}
\label{sec:sec2_3_3}
The pulsar's emission is so intense that the faint and extended PWN is not clearly visible in NuSTAR's 2D images. So we performed an analysis of the 5--50\,keV
radial profile of the NuSTAR data, as displayed in Figure~\ref{fig:fig3} (c).
Similar to the approach used for the Chandra and XMM-Newton profiles, we fit the NuSTAR profile with the PSF,
PWN function and background.

NuSTAR's PSF \citep[][]{amwb+14,mhma+15} is considerably broader compared to both the Chandra's PSF
and the Gaussian core of the PWN profile measured by Chandra.
As a result, the PSF-convolved PWN function essentially appears as
an exponential function. Besides, it is unclear whether the PWN profile in
the 5--50\,keV band is the same as the low-energy Chandra profile. In light of these complications, we opted for an exponential function instead of the PSF-convolved
PWN function for modeling the NuSTAR profile.
Upon analysis, the best-fit width of the exponential function was determined to be $l=28''\pm3''$, which is smaller than the Chandra-derived $l$ of $49.5''\pm5.8''$. This discrepancy possibly suggests a decrease in size with increasing energy, although it could also be attributed to the influence of the Gaussian core observed in the Chandra profile. 

\newcommand{\marka}{\tablenotemark{\tiny{\rm a}}}
\newcommand{\markb}{\tablenotemark{\tiny{\rm b}}}
\newcommand{\markc}{\tablenotemark{\tiny{\rm c}}}
\newcommand{\markd}{\tablenotemark{\tiny{\rm d}}}
\newcommand{\marke}{\tablenotemark{\tiny{\rm e}}}
\newcommand{\markf}{\tablenotemark{\tiny{\rm f}}}
\newcommand{\markg}{\tablenotemark{\tiny{\rm g}}}
\begin{table*}[t]
\vspace{-0.0in}
\begin{center}
\caption{Spectral analysis results}
\label{ta:ta2}
\vspace{-0.05in}
\scriptsize{
\begin{tabular}{lccccccccc} \hline\hline
Data       & {Instrument}\marka & Model\markb  & Energy & $N_{\rm H}$  & $\Gamma$/$a$ & $b$  &  $F_{\rm 2-10\,keV}$\markc & $\chi^2$/dof & Comment \\
           &   &     & (keV)   & ($10^{22}\rm \ cm^{-2}$) &  &  &  &   & \\ \hline
PSR        & XP+Ni+Nu & {\tt logpar} & 0.3--60 & $6.4\pm0.4$ & $1.40\pm0.03$ &$0.38\pm0.09$  &  $4.09\pm0.07$\markd    & 1325/1296 &  on$-$off \\
PSR        & XP+Ni+Nu & PL & 0.3--60 & $8.1\pm0.2$ & $1.42\pm0.03$ &$\cdots$  &  $4.29\pm0.07$\markd    & 1342/1297 &  on$-$off \\ \hline 
PSR        & XM+XP  & PL  & 0.3--10 & $6.7\pm0.2$ & $1.21\pm0.03$ &  $\cdots$ & $4.75\pm0.06$   & 820/811 & on+off   \\ 
PSR        & XP    & PL & 0.3--10 & 6.7\marke   & $1.20\pm0.04$ &  $\cdots$  & $3.85\pm0.06$\markd   & 456/410 &  on$-$off \\ \hline 
PWN        & CXO   & PL & 0.3--10 & 6.4\markf   & $1.96\pm0.33\pm0.12$\markg &  $\cdots$  & $1.44\pm0.18\pm0.02$\markg & 43/45 &   \\ 
PWN        & Nu  & PL & 5--20   & 6.4\markf   & $2.64\pm0.41\pm0.36$\markg &  $\cdots$  & $1.94\pm0.68\pm0.11$\markg     & 56/61   &  \\
PWN        & CXO+Nu   & PL & 0.3--20 & 6.4\markf   & $2.25\pm0.29\pm0.15$\markg &  $\cdots$  & $1.37\pm0.15\pm0.05\markg$ & 104/107 &          \\ \hline
\end{tabular}}
\end{center}
\vspace{-0.5 mm}
\footnotesize{
\marka{XP: XMM-Newton/PN, XM: XMM-Newton/MOS, Ni: NICER, Nu: NuSTAR, CXO: Chandra.}\\
\markb{$K(E/E_p)^{[-a-b \mathrm{log}(E/E_p)]}$ with $E_p=10$\,keV for {\tt logpar} and $K(E/1\rm \ keV)^{-\Gamma}$ for PL.}\\
\markc{Absorption-corrected 2--10\,keV flux in units of $10^{-12}\rm \ erg\ s^{-1}\ cm^{-2}$.}\\
\markd{Spin-cycle averaged flux.}\\
\marke{Fixed at the value obtained from the fit of the XM on+off spectrum.}\\
\markf{Fixed at the value obtained from the {\tt logpar} model of the XP+Ni+Nu on$-$off spectra.}\\
\markg{The second error denotes the systematic uncertainty (see Section~\ref{sec:sec2_4_3}).}\\
}
\end{table*}

The high temporal resolution of NuSTAR enabled us to independently measure the on- and off-pulse emissions. By analyzing the radial profiles of the on- and off-pulse data, we found that the off-pulse pulsar emission within the 5--50\,keV band is 22.1$\pm$1.5\% of the on$-$off emission. This estimation is consistent with the $22\pm1$\% obtained from the XMM-Newton/PN profile (Section~\ref{sec:sec2_3_1}). Consequently, we conclude that the ratio of counts in the off- and on-pulse emission does not vary significantly in the broad X-ray band. 

\subsection{Spectral analysis}
\label{sec:sec2_4}
As previously mentioned, the prominent pulsar emission dominates over
the PWN emission in the XMM-Newton and NuSTAR data.
This section aims to precisely characterize the pulsar's spectrum
by using the NICER, XMM-Newton, and NuSTAR data  (Section~\ref{sec:sec2_4_1}).
This is crucial to the NuSTAR measurement of the PWN spectrum because contamination of the PWN region by the pulsar's
emission is a concern in the NuSTAR analysis.
While the strong contamination from the on-pulse emission of the pulsar
can be minimized by using the off-pulse interval (Figure~\ref{fig:fig1}) in the measurement
of the PWN spectrum, there is still some contamination from the pulsar's off-pulse emission.
We characterize this off-pulse spectrum using the XMM-Newton data by comparing the ``on+off'' spectrum,
in which contamination from the PWN is negligible, with the ``on$-$off'' spectrum, as shown below (Section~\ref{sec:sec2_4_2}).
Additionally, we assess contamination by other point sources within the PWN (Section~\ref{sec:sec2_4_3}).
Subsequently, we take into account
emissions from the pulsar and other point sources when measuring the PWN spectra (Sections~\ref{sec:sec2_4_3} and \ref{sec:sec2_4_4}).

We performed spectral fitting in XSPEC v12.13.0c by employing the $\chi^2$ statistic.
To account for Galactic absorption, we adopted the {\tt tbabs} model with
the {\tt wilm} abundances \citep[][]{wam00} and the {\tt vern}
cross section \citep[][]{vfky96}.

\subsubsection{On$-$off spectrum of the pulsar J1849}
\label{sec:sec2_4_1}
To measure the ``on$-$off'' spectrum of J1849, we selected events within
the on- and off-pulse intervals (Figure~\ref{fig:fig1}) from the XMM-Newton/PN, NICER, and NuSTAR data. For extracting on-pulse (source) spectra, circular regions with radii of $R=16''$ and $R=60''$ were employed for XMM-Newton and NuSTAR, respectively.
The off-pulse (background) spectra were extracted within the same regions
used for the source regions.
As for the NICER data, source and background spectra were extracted from each of the 108 observations. We then combined these NICER spectra using the {\tt addascaspec} script to construct high-quality source and background spectra.
Corresponding spectral response files were generated
following the standard procedure suitable for each observatory.
We then grouped the source spectra to ensure at least 20 counts per spectral bin.

We jointly fit the on$-$off spectra measured by XMM-Newton, NICER, and NuSTAR with an absorbed power-law (PL) or
the log parabolic power-law \citep[{\tt logpar};][]{Massaro2004} model.
For each spectrum, a cross-normalization factor was incorporated, with the factor for the NICER spectrum set to 1.
The energy bands used were 0.3--10\,keV, 2--8\,keV, and 5--60\,keV for the XMM-Newton, NICER, and NuSTAR spectra, respectively. The selection of these energy bands aimed to minimize contamination from background and (cross-) calibration uncertainties \citep[e.g.,][]{Madsen2022}.

\begin{figure*}
\centering
\begin{tabular}{ccc}
\hspace{-2mm}
\includegraphics[width=2.5 in]{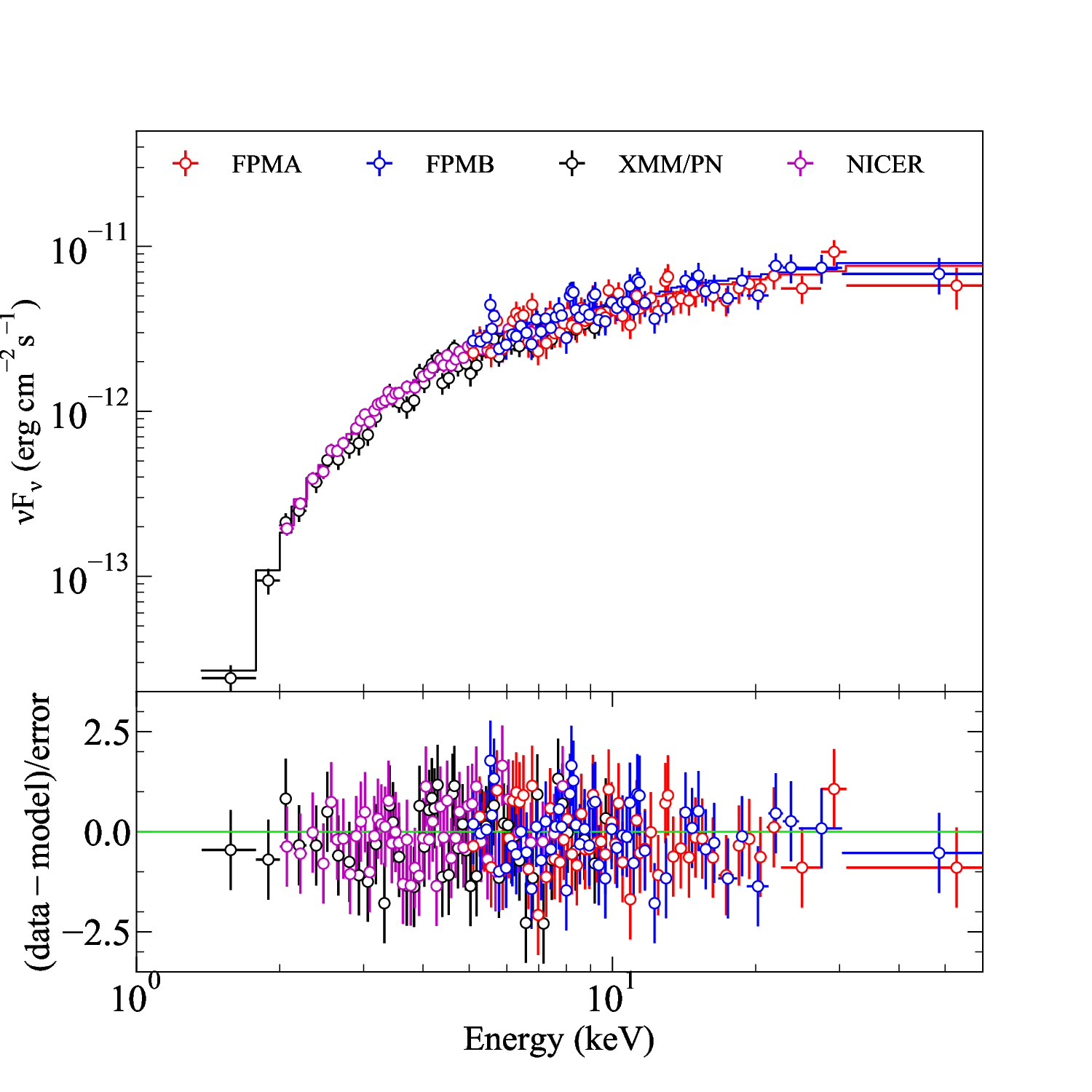} &
\hspace{-8mm}
\includegraphics[width=2.5 in]{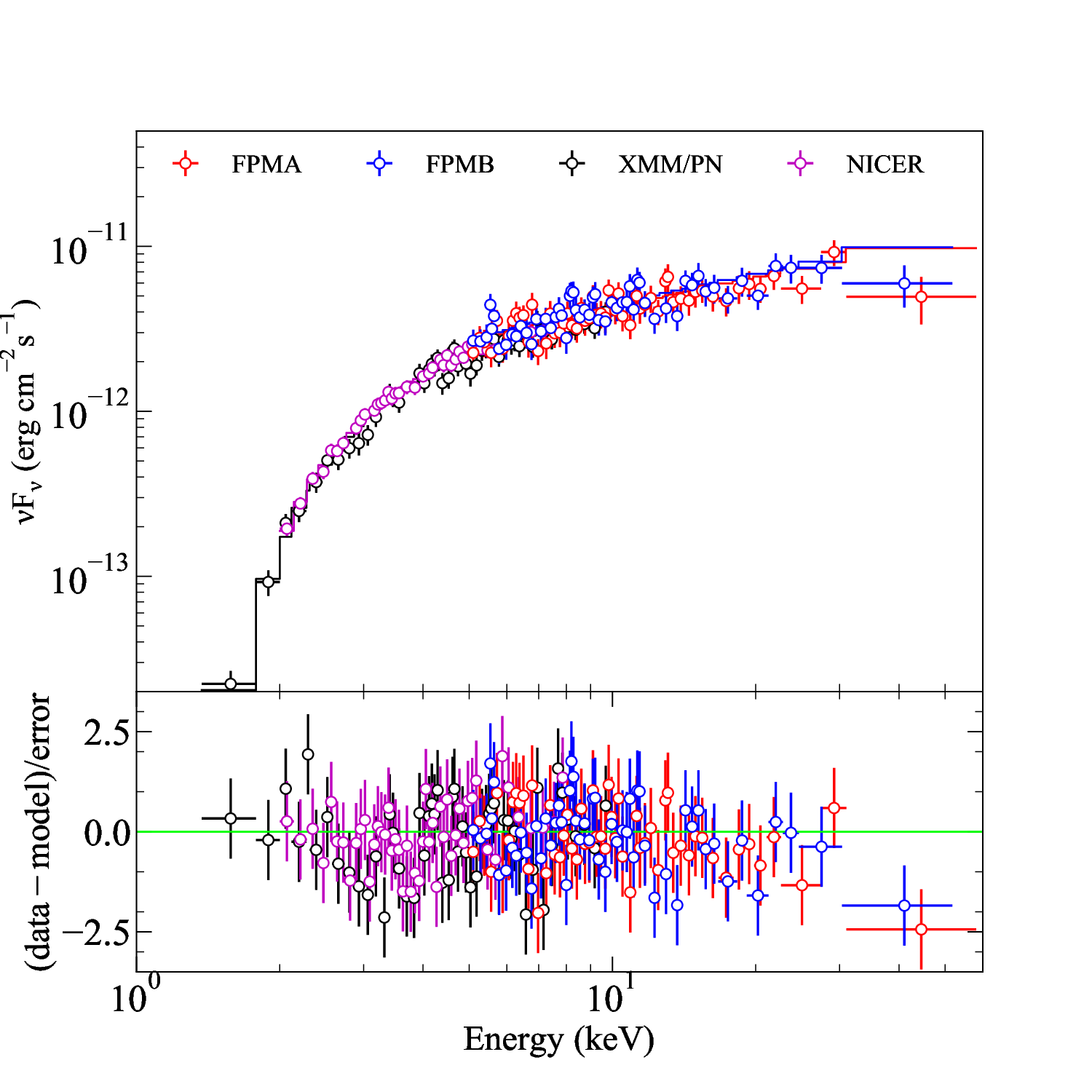} &
\hspace{-8mm}
\includegraphics[width=2.5 in]{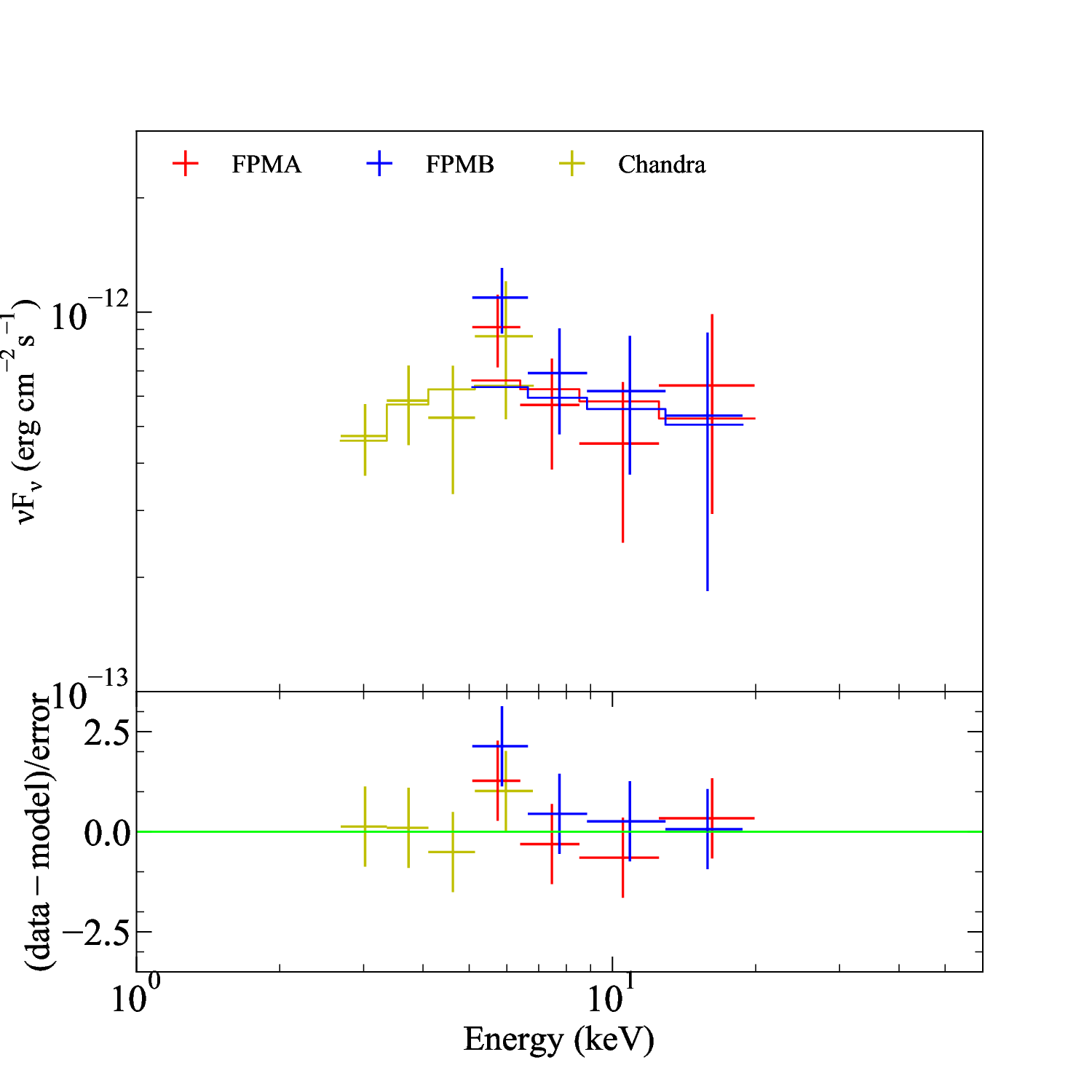} \\
\end{tabular}
\figcaption{({\it Left}): X-ray spectra of J1849 measured with XMM-Newton (black), NICER (purple), and NuSTAR (red and blue) and the best-fit {\tt logpar} model (solid curves). The pulsar spectra were measured in the on-pulse interval by subtracting the off-pulse background (i.e., on$-$off) and were averaged over a spin cycle. ({\it Middle}): PL fit of the on$-$off spectra of J1849. ({\it Right}): The PWN spectra measured with Chandra (yellow) and NuSTAR (red and blue) and the best-fit PL model.
The bottom panels show the fit residuals
\label{fig:fig4}}
\vspace{0mm}
\end{figure*}

While the PL fit was acceptable, adopting the {\tt logpar} model resulted in an improved fit (Table~\ref{ta:ta2} and Figure~\ref{fig:fig4}).
The latter model was favored with an $f$-test probability of $5\times 10^{-5}$.
The cross-normalization factors for the NuSTAR spectra were consistent with
1 at $\lapp1\sigma$ levels. However, the factor for the XMM-Newton/PN spectrum was measured to be  0.94$\pm$0.02, exhibiting a deviation from 1 at a $\sim3\sigma$ significance level. This discrepancy might indicate some systematic effects, potentially stemming from cross-calibration issues \citep[see][]{Madsen2017}. These cross-normalization factors are taken into account below while measuring the PWN spectrum.

Our estimation of $N_{\rm H}=(6.4\pm0.4)\times 10^{22}\rm \ cm^{-2}$ towards the source
is larger than previous results of (4.0--4.5)$\times 10^{22}\rm \ cm^{-2}$ \citep[][]{Terrier2008,Gotthelf2011,kh15,Vleeschower2018}.
We believe that the discrepancy could be ascribed to a different abundance model that these authors used.
Although they did not report the abundance model, we were able to reproduce their $N_{\rm H}$ values
by using the {\tt angr} abundances (default in XSPEC).

\subsubsection{Estimations of the off-pulse spectrum of the pulsar}
\label{sec:sec2_4_2}
Estimating the pulsar emission within the `off-pulse interval' is crucial
for accurately measuring the $R=150''$ PWN spectrum using the NuSTAR data. Based on our analysis results presented so far, we considered that 
contamination of the PWN region by the pulsar's off-pulse emission would not be negligible.
While our image analysis already suggests that the off-pulse spectral shape is similar to that of the `on$-$off' spectrum, XMM-Newton's moderate angular resolution offers another means to probe the off-pulse spectral shape. By focusing on an $R=16''$ circular region centered at the pulsar, we found that the PWN contributes less than 2\% to the measured XMM-Newton on-pulse counts (Section~\ref{sec:sec2_3_1}). This contribution can be further reduced by adequately selecting a background region and would be so minor that we can safely assume the spectrum of the $R\le16''$ region accurately represents the pulsar's emission. This assumption enables us to estimate the ``off-pulse'' spectrum through a comparison of the ``on+off'' and ``on$-$off'' spectra. For this comparison, we employ a simple PL model.

We analyzed the XMM-Newton (MOS and PN) data to measure the on+off spectra within $R=16''$ circles
centered at the pulsar. Background spectra were extracted from source-free regions on the same detector chips as the source regions, employing $R=32''$ circles.
We fit the spectra with a PL model, allowing for a variable cross-normalization factor for each instrument. These factors were found to be consistent with 1 at $\lapp 1\sigma$ confidence levels.
The results are presented in Table~\ref{ta:ta2}.

Although we have measured the ``on$-$off'' spectrum of the pulsar using the multi-mission data in the previous section,
we fit the XMM-Newton/PN ``on$-$off'' spectrum with a PL model for a comparison with the ``on+off'' spectrum measured with MOS+PN.
The results are presented in Table~\ref{ta:ta2}.
The best-fit parameter values for this ``on$-$off'' spectrum are in complete agreement with
those derived from the ``on+off'' spectrum (Table~\ref{ta:ta2}).
By comparing the on$-$off and on+off spectra, we find that the off-pulse flux is $\approx$23\% of the
on$-$off flux, which is in accord with the finding in our image analysis (Section~\ref{sec:sec2_3}),
where we determined that the off-pulse counts are $22\pm1$\% of the ``on$-$off'' counts.

We further verified these results by fitting the on+off spectra (MOS1,2 and PN) with two PL components.
The first PL component corresponds to the `off-pulse' emission, while the second one describes the on$-$off spectrum.
The parameters of the second PL component were held fixed at the values derived from the on$-$off spectrum
measured with the XMM-Newton/PN data. The best-fit $\Gamma$ for the first component is $1.22\pm0.09$,
and the absorption-corrected 2--10\,keV flux ($F_{\rm 2-10 keV}$)
was determined to be $(9.1\pm0.5)\times 10^{-13}$\,\fluxcgs,  which is
consistent with the above results. In summary, we found out that the off-pulse flux is $\approx$22\% of the on$-$off flux and the off-pulse emission exhibits a very similar spectral shape to the on$-$off spectrum.

\subsubsection{Spatially-integrated spectrum of the PWN~G32.6}
\label{sec:sec2_4_3} 
To characterize the spatially-integrated emission of the large PWN efficiently,
it is necessary to reduce contamination from the pulsar through spatial or temporal selection.
The XMM-Newton data do not allow for such selection; therefore,
we measure the broadband X-ray spectra of the $R=150''$ PWN using the Chandra and NuSTAR data.
In this case, in-flight backgrounds may not accurately represent those
in the large source region because of various effects, such as
inhomogeneity in the detector background and optics vignetting.
The sky (photon) background is affected by optics vignetting,
whereas some background components, such as particle-induced background, are not affected
by the optics. We remove the latter by subtracting the blank-sky data and measure the former
using the blank-sky-subtracted in-flight spectrum taken in off-source regions
for the Chandra analysis (see below).
For the NuSTAR data, we use the {\tt nuskybgd} simulations \citep[][]{Wik2014}.
More detailed explanations of these background estimation processes can be found in \citet{Park2023a}.

For the Chandra data, we generated blank-sky events appropriate for the observational data using
the {\tt blanksky} script of CIAO.
We extracted ACIS spectra from
both the observational and blank-sky data within an annular region centered at J1849 with the inner and outer radii of $5''$ and $150''$, respectively. We minimized contamination from the point sources S1--11 (Figure~\ref{fig:fig2} (a)) by excising an $R=2''$ circular region centered at each source.
We then subtracted the blank-sky spectrum from the observed spectrum to construct
a source spectrum. The same procedure (i.e., subtraction of the blank-sky data) was applied to extract background spectra from
regions located away (5$'$--6$'$) from the source region.
To further refine the analysis, we corrected the blank-sky-subtracted background spectrum for the vignetting effect by multiplying it with the ratio of the energy-dependent effective areas (i.e., ancillary response files) corresponding to the source and background regions.

We fit the source spectrum using a PL+{\tt logpar} model, where
the {\tt logpar} model accounts for the contamination by the pulsar.
To model the pulsar emission (i.e., {\tt logpar}), we adopted point-source response files corresponding
to the annular source region ($R=5''$--$150''$), which took into account the reduced enclosed energy fraction of
Chandra's PSF. The parameters of the {\tt logpar} model were set to
the values obtained from the on$-$off spectrum (Section~\ref{sec:sec2_4_1}), and we
introduced a normalization factor to account for the
22\% off-pulse emission. These pulsar parameters were then held fixed.
By optimizing $N_{\rm H}$ of the PWN model independently of the value for the pulsar model, we determined
$N_{\rm H}=(6.6\pm3.6)\times 10^{22}\rm \ cm^{-2}$, which is consistent with that
for the pulsar model. Therefore, we linked $N_{\rm H}$ of the PL model to that of the {\tt logpar} model.
The PL+{\tt logpar} model was acceptable, with
the best-fit PL parameter values reported in Table~\ref{ta:ta2}.

The point sources S1--11 (Figure~\ref{fig:fig2} (a)) may contaminate the NuSTAR data in the large $R=150''$ region. 
We therefore measured their summed spectrum using the Chandra data. We constructed a source spectrum using $R=2''$
circular regions centered at each source. A background spectrum was extracted within $R=3''$ circular regions near the source
regions. We then fit the spectrum with a PL model and inferred the best-fit parameters to be
$N_{\rm H}=(5.8\pm2.1)\times 10^{22}\rm \ cm^{-2}$, $\Gamma=1.35\pm0.66$, and $F_{\rm 2-10\ keV}=(2.2\pm0.3)\times 10^{-13}$\,\fluxcgs. When these point sources were not excised from the Chandra data, the inferred PWN flux increased by $\approx$15\%, but the photon index changed only by $\approx$0.03. Their influence on the NuSTAR data could be larger since the spectrum is hard. While it is unclear whether the hard spectrum extends to $>$8\,keV, we use this PL model to account for the contamination in the NuSTAR analysis as a conservative estimate.

We extracted NuSTAR spectra of the PWN utilizing the off-pulse interval to
minimize the contamination by the pulsar (pulse gating). We generated the source spectra
using an $R=150''$ circular region and
estimated the background in the source region using the {\tt nuskybgd} tool \citep[][]{Wik2014}.
We ensured that the simulated background spectrum matched the source-region spectrum well
at $>$30\,keV, where background emission is expected to be dominant.
We fit a PL+{\tt logpar}+PL model to the NuSTAR spectra.
As in the case of the Chandra analysis (see above), the {\tt logpar} model accounts for
the off-pulse emission. For this model, we used point-source response files and adopted the
{\tt logpar} parameters listed in Table~\ref{ta:ta2}, along with an additional normalization factor suitable for the lower flux of the off-pulse emission (22\%);
these pulsar parameters were frozen.
The second PL model was introduced to take into account the contamination from S1--11; its parameters were
held fixed at the Chandra-inferred values.
The best-fit parameters for the first PL model (i.e., PWN emission) are presented in
Table~\ref{ta:ta2}. At face value, the NuSTAR spectra are softer than the Chandra spectrum, possibly indicating a spectral break. However, the difference in the best-fit $\Gamma$ values is statistically insignificant.
Moreover, modeling the pulsar and S1--11 may introduce systematic uncertainties (see below).

Next, we jointly fit the Chandra and NuSTAR spectra using the PL+{\tt logpar}+PL model,
incorporating a cross-normalization factor for each instrument.
Because the contamination from the point sources was excised from the Chandra data, the flux of the second PL was set to 0 for the Chandra spectrum. We held the parameters of the second PL component (emission from S1--S11) in the NuSTAR model fixed at the Chandra-inferred values.
The model adequately fit the spectra (Figure~\ref{fig:fig4} right), and the best-fit parameter
values are presented in Table~\ref{ta:ta2}. The cross-normalization factors
of the two instruments agreed within the $\le1\sigma$ level.
Additionally, we fit the spectra with a broken-PL (BPL) plus {\tt logpar} and PL.  The best-fit parameters of the BPL model are $\Gamma_l=1.8\pm 0.3$, $\Gamma_h=3.4\pm 0.9$, and $E_{\rm brk}=6 \pm 2$\,keV, but the BPL and PL models are statistically indiscernible with an $f$-test probability of 0.2.

Since the results may vary depending on the background selection, the assumed pulsar ({\tt logpar} parameters) and point-source (PL parameters for S1--S11; relevant to NuSTAR) models,
we evaluated systematic uncertainties in the inferred PL parameters for the PWN emission.
We performed spectral analysis using various background estimations (e.g., regions), and {\tt logpar} and point-source model parameter values
(within their uncertainties). We then quantified the 1$\sigma$ changes of the PWN model parameters,
which are presented as the additional uncertainty in Table~\ref{ta:ta2}.

\begin{figure*}
\centering
\begin{tabular}{ccc}
\hspace{-3mm}
\includegraphics[width=2.4 in]{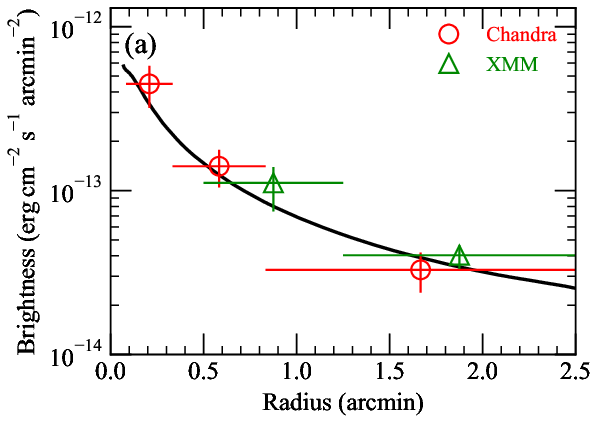} &
\hspace{-5mm}
\includegraphics[width=2.3 in]{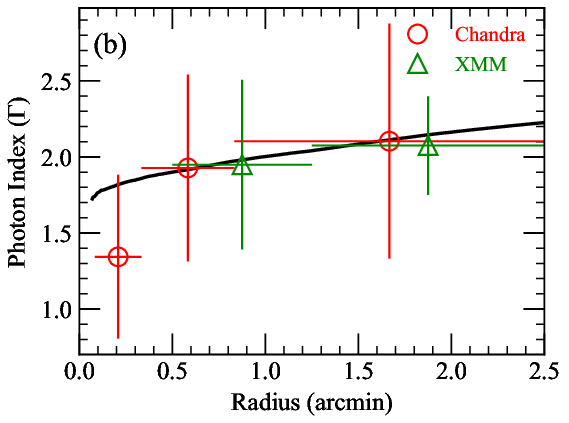} &
\hspace{-5mm}
\includegraphics[width=2.4 in]{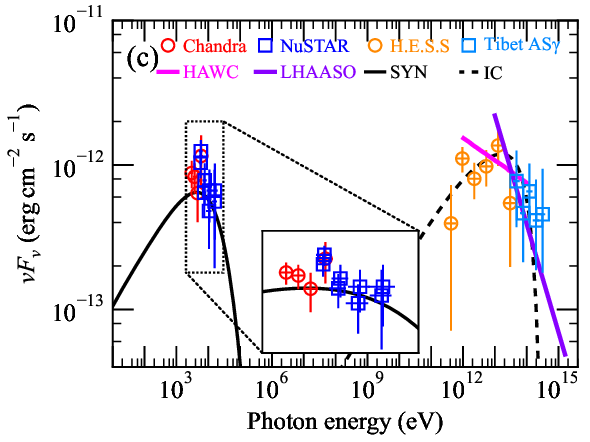} \\
\end{tabular}
\figcaption{Radial profiles of the 2--10\,keV brightness and $\Gamma$, and a broadband SED of G32.6.
(a--b) Red and green points denote Chandra and XMM-Newton measurements, while black curves represent
radial profiles computed by our model. (c) Red and blue points are X-ray measurements
obtained from Chandra and NuSTAR, respectively. The empty orange circles 
in the TeV band correspond to H.E.S.S. measurements taken from \citet[][]{HESSHGPS2018},and
the cyan squares display Tibet AS$\gamma$ measurements \citep[][]{Amenomori2023}.
 Straight lines in the TeV band represent the best-fit PL models inferred from the HAWC \citep[pink;][]{HAWC2022} and LHAASO \citep[purple;][]{LHAASO2023} data.
Our model computations for the synchrotron (SYN) and inverse-Compton (IC) emissions are
presented in black solid and dashed curves, respectively.
\label{fig:fig5}}
\vspace{0mm}
\end{figure*}

\subsubsection{Spatially-resolved spectra of the PWN}
\label{sec:sec2_4_4}
We performed a spatially-resolved spectral analysis of the Chandra data using
three annular regions with sizes of $R=5''$--$20''$, $20''$--$50''$ and $R=50''$--$150''$,
to investigate possible spectral variation within the PWN.
We extracted ACIS spectra from these regions, and fit them along with
background spectra constructed from the blank-sky data (Section~\ref{sec:sec2_4_2}).
We jointly fit the spectra, taking into account the contamination by the pulsar
(i.e., {\tt logpar} model; see Section~\ref{sec:sec2_4_3}).
Our analysis revealed an increase in $\Gamma$ with increasing distance from the pulsar (Figure~\ref{fig:fig5});
however, this change is statistically insignificant due to large uncertainties.

We also used the XMM-Newton/MOS data to investigate the spatial variation of the PWN spectrum.
We extracted spectra from annular regions of $R=30''$--$75''$ and $R=75''$--$150''$, excluding contaminating sources within the regions. For background estimation, we followed the same procedure employed in the Chandra
analysis (Section~\ref{sec:sec2_4_3}), using the quiescence particle background (qpb)
data generated by {\tt evqpb} of SAS, instead of the blank-sky data used for Chandra \citep[see][for more detail]{Park2023a}.
We then fit the spectra with the PL+{\tt logpar} model, where the {\tt logpar} model accounts for the contamination from the pulsar. As in the Chandra analysis, we introduced a normalization factor
to accommodate the off-pulse emission and employed point-source response functions for
the {\tt logpar} model. The PL+{\tt logpar} model provided a satisfactory fit to the data and the best-fit
parameters for the PL component are presented in Figure~\ref{fig:fig5}.

These XMM-Newton measurements are consistent with the Chandra results, albeit yielding a larger $\Gamma$ ($\approx 2.1$) for the $R=75''$--$150''$ region compared to previous analyses of the same region \citep[$\Gamma$=1.7--1.8;][]{Terrier2008,kh15,Vleeschower2018}. Although the difference is statistically insignificant, given the substantial uncertainties, we speculate that the contamination from the pulsar emission in the previous analyses may have contributed to the discrepancy. By omitting the pulsar model from our analysis, we were able to reproduce the previous results.
It is worth noting that \citet{Gotthelf2011} also considered the contamination by the pulsar and
inferred $\Gamma=2.1\pm0.3$ for the PWN spectrum within an $R=30''$--$150''$ annular region.

We also estimated systematic uncertainties for these spatially-resolved spectra as we did in Section~\ref{sec:sec2_4_3}.
The systematic uncertainties were added to the statistical ones in quadrature, and Figure~\ref{fig:fig5} displays the summed uncertainties.

\section{Modeling the PWN emission}
\label{sec:sec3}
We constructed a broadband SED of the PWN from X-rays to TeV energies, by supplementing
our X-ray measurements with published TeV data \citep[][]{HESSHGPS2018,3HWC2020,LHAASO2023}.
The resulting broadband SED is presented in Figure~\ref{fig:fig5} (c). 
Our X-ray characterization of the PWN emission has improved upon
previous studies \citep[e.g.,][]{Terrier2008, Gotthelf2011, kh15, Vleeschower2018}, as we
included the $>$10\,keV NuSTAR data and carefully accounted for contamination
from the pulsar and S1--S11.
We employed the multizone PWN model developed by \citet{KimS2019} to fit the SED as well as radial profiles of the X-ray brightness and $\Gamma$.

\subsection{Description of the multizone model}
Since our multizone PWN model is extensively described in \citet{Park2023a}, we provide a concise overview hereafter. 
We assume that the pulsar injects electrons  with a power-law distribution
\begin{equation}
\label{eq:edist}
\frac{dN_e}{d\gamma_e dt}=\dot N_0 \gamma_e^{-p_1}
\end{equation}
between
$\gamma_{e,\rm min}$ and $\gamma_{e,\rm max}$ into the TS at a distance of $R_{\rm TS}$ from the pulsar such that the particle energy corresponds to a fraction $\eta_e$ of the pulsar's spin-down power.
The remaining power is contained in $B$ (fraction of $\eta_B$) or radiated as gamma-ray emission from the pulsar (fraction of $\eta_\gamma$).
As the magnetic energy is thought to be almost fully dissipated at the shock \citep[][]{kc84b}, we assume $\eta_B\ll1$ in the PWN. Additionally, the pulsar (J1849) was not detected by the Fermi Large Area Telescope
\citep[LAT;][]{Atwood2009}, indicating $\eta_\gamma\ll 1$.
Therefore, in the following study, we assume $\eta_e\approx 1$.
These parameters affect the energy of the electrons injected by the pulsar into the PWN,
and thus the predicted flux of
the PWN could vary depending on the assumed $\eta_e$.
A smaller value of $\eta_e$ (i.e., larger $\eta_\gamma$) would
cause the model to underpredict the measured fluxes (but see Section~\ref{sec:sec3_2}).

The electrons propagate within the PWN via advection and diffusion, and they lose energy
due to radiation (synchrotron and inverse-Compton) and adiabatic expansion.
Regarding the bulk flow ($V_{\rm flow})$
of the electrons and $B$ within the PWN, we adopt a power-law dependence on the
distance ($r$) between the pulsar and an emission zone:
\begin{equation}
\label{eq:vflow}
V_{\rm flow}(r)=V_0 \left ( \frac{r}{R_{\rm TS}} \right )^{\alpha_V}
\end{equation}
and
\begin{equation}
\label{eq:bfield}
B(r)=B_0 \left ( \frac{r}{R_{\rm TS}} \right )^{\alpha_B},
\end{equation}
where we assume $\alpha_B+\alpha_V=-1$, which holds if the $B$ structure
is toroidal and the magnetic flux is conserved \citep[e.g.,][]{Reynolds09}.
We assume the diffusion coefficient to be \citep[e.g.,][]{HAWC_halo2017}
\begin{equation}
\label{eq:diffusion}
D=D_0 \left ( \frac{B}{100\mu\rm G} \right )^{-1} \left ( \frac{\gamma_e}{10^9} \right )^{1/3}.
\end{equation}

We let the injected electrons propagate over an assumed age of the PWN,
and then we project their radiation onto the tangent plane of the observer to compute both spatially-resolved and integrated SEDs.

\subsection{Basic model parameters}
\label{sec:sec3_1}
While the model employs multiple parameters, some of them can be
estimated first based on the SED measurements and one-zone model calculations. These estimations serve as inputs for the detailed modeling (Section~\ref{sec:sec3_2}) and help with obtaining converged solutions more quickly. 

The similar values of $\nu F_\nu$ for the X-ray (synchrotron) and TeV (gamma-ray) emission indicate that
the magnetic energy density ($U_B\equiv \frac{B^2}{8\pi}$) is comparable to the energy density
of the CMB ($u_{\rm CMB}$) seed photons for inverse-Compton (IC) upscattering.
From this, we estimated $B\sim 3\mu$\,G.
Other complications, such as the contribution of the Galactic IR field and spatially varying $B$, are
considered by the multizone model.
Additionally, the measured TeV-to-X-ray flux ratio can provide information on the true age of the PWN. According to the correlations reported by \citet{Zhu2018},
the flux ratio of $\sim$1 for G32.6 suggests that its true age is 4--9\,kyr, much
smaller than the spin-down age ($\tau_c= 43$ kyr) of J1849.
We assume $t_{\rm age}$ of 9\,kyr in this work, but models with different values of
$t_{\rm age}$ (and other covarying parameters, e.g., $D_0$) can also fit the data well
\citep[see][for parameter covariance]{Park2023a}.

Given the estimated $B$ strength ($B\approx 3\mu$G), the synchrotron emission at $\sim$20\,keV detected by NuSTAR suggests
$\gamma_{e,\rm max}$ of $\ge 6\times 10^8$.
The X-ray photon index of $\Gamma\approx 1.5$ in the innermost region corresponds
to the power-law index of the uncooled particle distribution of $p_1 =2\Gamma -1 \approx 2$.
These estimations of $\gamma_{e,\rm max}$ and $p_1$
allow for the determination of $\gamma_{e,\rm min}$ since $\eta_e\approx 1$.

Although the TS of G32.6 has not been resolved likely due to its compact size obscured by the bright pulsar emission, we assumed $R_{\rm TS} = 0.1$\,pc as has been measured for several PNWe \citep[e.g.,][]{nr04,Kargaltsev2008}.
This assumption is also in accord with the correlation reported by \citet{Bamba2010b}.
Note that the exact value of $R_{\rm TS}$
does not significantly impact the model outputs as long as it is sufficiently smaller than the PWN size.
Although the X-ray PWN does not exhibit a sharp boundary, we assumed 
an X-ray PWN size of $R_{\rm PWN}=150''$; this radial position is where we
halt the advection flow and lower $B$ to a slightly smaller value (2.3$\mu$G in the PWN to 2$\mu$G outside).
 When $r>150''$ (outside the X-ray PWN region), we do not invoke $\alpha_B+\alpha_V=-1$;
both $B$ and $V_{\rm flow}$ are assumed to be constant.
The large region for the TeV emission ($R=0.09^\circ$) corresponds to 11\,pc for an assumed distance of 7\,kpc \citep[][]{HESSHGPS2018}
and implies that the TeV-emitting electrons ($\gamma_e\approx 10^7$) can go beyond the X-ray PWN due to diffusion. 
Considering the diffusion only, $2\sqrt{D t_{\rm age}} \gapp 10$\,pc gives an estimate of $D_0 \gapp 1\times 10^{26}\rm \ cm^2\ s^{-1}$ for the assumed age of $t_{\rm age}=9$\,kyr \citep[e.g., see][for values of diffusion coefficients for various PWNe]{Tang2012,Porth2016,Guest2019}.

\subsection{Results of modeling}
\label{sec:sec3_2}
We used the aforementioned estimations of the parameters (Section~\ref{sec:sec3_1})
as initial values for our multi-zone model
and further adjusted them to fit the measured emission properties of G32.6.
For the seeds of the IC emission, we adopted the interstellar radiation field (ISRF) estimated by {\tt GALPROP} \citep[][]{Porter2022},
along with the CMB radiation.
The results are presented in Figure~\ref{fig:fig5} and Table~\ref{ta:ta3}. 
Our model broadly matches the broadband SED, as well as the radial profiles of brightness and $\Gamma$. 

The amplitude of the model-predicted SED is determined by a complex interplay
between the spectrum of the injected electrons (e.g., $\eta_e$ and $p_1$),
the time the electrons spend in the emission region (e.g., $t_{\rm age}$ and $V_{\rm flow}$),
and the environmental factors ($B$ and external IR seed density $u_{\rm IR}$).
In this work, we assumed a large value of $\eta_e$, indicating weak gamma-ray emission from the pulsar,
since J1849 was not detected by the LAT.
We verified that different values of $\eta_e$ could be easily accommodated in our model
by adjusting other parameters
such as $t_{\rm age}$, $B$, $u_{\rm IR}$, and/or $p_1$ \citep[see][for the parameter covariance]{Park2023a}.

An electron injection spectrum with $p_1\approx 2$ was necessary to match the X-ray SED
as well as the $<$10\,TeV data. For this $p_1$ value, a $\Gamma=1.5$ PL emission from uncooled electrons is expected. These uncooled electrons are responsible for the $\Gamma\approx 1.5$ X-ray spectrum in the inner $R\lapp 20''$ region. On the other hand, the $\Gamma\approx 2$ spectra in the outer regions are produced by cooled electrons. This implies that the cooling break (i.e., the peak of the curved SED) should be visible in the spatially-integrated X-ray SED. 
The computed SED model predicts a peak at 5\,keV. Hence, the model predicts that the $<$5\,keV SED exhibits a harder spectrum than above the energy, as indicated by the separate fits of the Chandra and NuSTAR data; however, the broadband SED fits well with a single PL probably because of the large measurement uncertainties.

In this model, TeV emission is primarily produced by IC scattering
of electrons with $\gamma_e\approx 10^7$--$10^8$ off of the CMB photons.
Given the inferred $B$ value, the assumed $t_{\rm age}$ of 9\,kyr corresponds to
the synchrotron+IC cooling timescale 
of $\gamma_e\sim 8\times10^7$ electrons.
Consequently, a spectral break of the IC SED is expected at $\sim$40\,TeV.
Moreover, the Klein-Nishina effect becomes significant for $\ge 20$\,TeV (IR seeds) and $\ge 100$\,TeV (CMB seeds) gamma rays. These factors explain the steep LHAASO SED at $\gapp$10\,TeV.

The radial profiles of brightness and $\Gamma$ provide constraints on $\alpha_B$ (and thus $\alpha_V=-1-\alpha_B$) and $D_0$ values. We used a slowly varying $B$, but different sets of $\alpha_B$ and $D_0$ can also reproduce the profiles.
For the measured radial profiles and the assumed power-law trend for $V_{\rm flow}$ (Equation~(\ref{eq:vflow})), our model constrains $\alpha_V$ to be between $-0.8$ and $-0.6$.
However, it is important to emphasize that the choice of a power law for $V_{\rm flow}$ was made on a phenomenological basis and lacks a physical motivation
and that the actual functional form of the flow in PWNe is not well-known. Our model has the flexibility to accommodate various trends; e.g.,
the radial profiles of $V_{\rm flow}$ and $B$ predicted for ideal magnetohydrodynamic flow by \citet{kc84a} \citep[see also][]{Reynolds2003} can also fit the data when we freely adjust the other parameters (e.g., $V_0$, $B_0$ and $D_0$).
These parameter values cannot be determined uniquely because of the covariance between them. The degeneracy can be broken to a certain extent if we can constrain some of the parameters; e.g., measuring the expansion speed of the PWN can help determine $V_0$ and $\alpha_V$ (Equation~(\ref{eq:vflow})), and $B_0$ can be well constrained if $u_{\rm CMB}+u_{\rm IR}$ are known. Nonetheless, the values presented in Table~\ref{ta:ta3}, which are similar to values inferred for other PWNe \citep[e.g.,][]{Park2023b}, can account for the measurements.

\begin{table}[t]
\vspace{-0.0in}
\begin{center}
\caption{Parameters for the SED model in Figure~\ref{fig:fig5}}
\label{ta:ta3}
\vspace{-0.05in}
\scriptsize{
\begin{tabular}{lcc} \hline\hline
Parameter  & Symbol   & Value   \\ \hline
Spin-down power       & $L_{\rm SD}$              & $9.8\times{10}^{36}\rm \ erg\ s^{-1}$     \\
Characteristic age of the pulsar & $\tau_{\rm c}$   & 43\,kyr        \\
Age of the PWN        & $t_{\rm age}$              &  9000\,yr        \\
Size of the PWN       & $R_{\rm pwn}$  & 5\,pc        \\
Radius of termination shock & $R_{\rm TS}$  & 0.1\,pc        \\
Distance to the PWN  & $d$  & 7.0\,kpc        \\ \hline
Index for the particle distribution   & $p_1$            & 2.12          \\
Minimum Lorentz factor  & $\gamma_{e,\rm min}$  & $10$        \\
Maximum Lorentz factor  & $\gamma_{e,\rm max}$ & $10^{8.9}$        \\
Magnetic field        & $B_0$      & 7.5$\mu$G        \\
Magnetic index        & $\alpha_B$   & $-$0.3 \\
Flow speed            & $V_0$        & 0.04$c$ \\
Speed index           & $\alpha_V$    & $-$0.7 \\
Diffusion coefficient & $D_0$    & $1.0\times 10^{26}\rm \ cm^2 \ s^{-1}$ \\
Energy fraction injected into particles   &  $\eta_e$      &  0.95        \\
Energy fraction injected into $B$ field        & $\eta_{B}$    & 0.004      \\ \hline
\end{tabular}}
\end{center}
\vspace{-0.5 mm}
\end{table}

\section{Discussion}
\label{sec:sec4}
We conducted a detailed X-ray analysis of both the pulsar and PWN of the J1849+G32.6 system. We found out that the on$-$off emission of the pulsar exhibits a curved spectrum
and the off-pulse emission has a similar spectral shape to the on$-$off one.
By carefully accounting for the contamination from the pulsar's emission in the low-imaging resolution XMM-Newton and NuSTAR data, we were able to characterize the PWN emission up to $\sim$20\,keV.
The PWN emission was modeled with a simple PL having $\Gamma\approx 2$. 
Our results from a spatially-resolved analysis are consistent with
the previously suggested spectral softening in G32.6 \citep[][]{kh15}, and we
measured the radial profiles of $\Gamma$ and brightness. Combining our
X-ray measurements with TeV SEDs, we modeled the broadband data using a
multi-zone emission model for PWNe.

\subsection{The Pulsar J1849}
\label{sec:sec4_1}
A curvature in the on$-$off spectrum of J1849 has been suggested by \citet{Terrier2008} and \citet{kh15} based on comparisons of a low-energy on$-$off spectrum and a high-energy on+off spectrum. While these comparisons hinted at the presence of curvature, the definite establishment was hindered as the comparisons involved distinct quantities, on$-$off vs. on+off emissions. Furthermore, issues such as contamination from the PWN in the INTEGRAL data and cross-calibration problems were not addressed in the previous studies. By combining the high-quality X-ray data acquired by XMM-Newton, NICER, and NuSTAR, we were able to conclusively demonstrate the evidence of a spectral curvature in the on$-$off pulse emission of J1849.

The pulsar J1849 is among the four `MeV pulsars'
listed in \citet{Harding2017} \citep[see][for more sources]{kh15}.
These pulsars are young and energetic rotation-powered pulsars (RPPs) that
exhibit strong emission of pulsed non-thermal X-rays while lacking
significant radio or GeV emission.
\citet{Harding2017} proposed that accurately characterizing
the SEDs of MeV pulsars can aid in understanding the mechanism of
pair production in RPPs. According to their study, the frequency of the SED peak ($E_{\rm SR}$) of the pulsar emission is
predicted to be $\propto B_S B_{\rm LC}$, where $B_{\rm LC}$ represents $B$
at the light cylinder, if the pairs are generated by polar-cap
cascades. On the other hand, if the pairs are produced in the
outer gap, $E_{\rm SR}\propto B_{\rm LC}^{7/2}$ is expected \citep[][]{Harding2017,Zhang2000}.

\begin{figure}
\centering
\includegraphics[width=3.05 in]{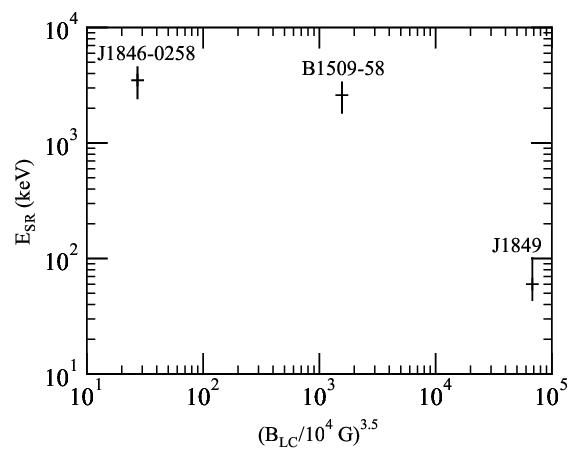} \\
\includegraphics[width=3.00 in]{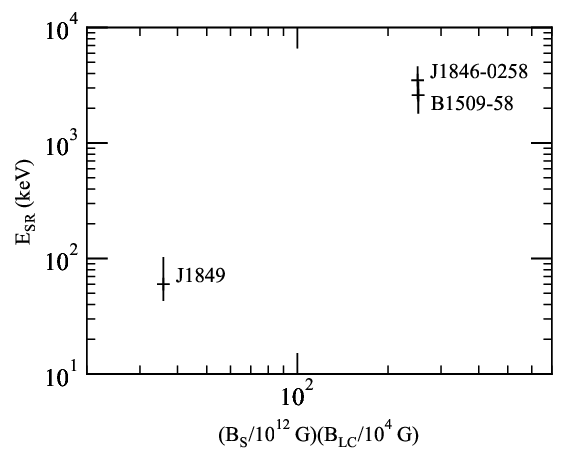} \\
\figcaption{Measured SED peak frequencies ($E_{\rm SR}$) of the synchrotron emission from three MeV pulsars
versus $B$ at the light cylinder ($B_{\rm LC}$; top) or a combination of $B_s$ and $B_{LC}$ (bottom).
\label{fig:fig6}}
\vspace{0mm}
\end{figure}

Measurements of the SED peaks for
PSR~B1509$-$58 \citep[$E_{\rm SR}=2.6\pm0.4$\,MeV;][]{Chen2016}
and PSR~J1846$-$0257 \citep[$E_{\rm SR}=3.5\pm1.1$\,MeV;][]{Kuiper2018}
indicate that the outer-gap scenario is improbable. This is evident as these two pulsars
possess substantially different $B_{\rm LC}$ values, yet
their $E_{\rm SR}$ values are very similar (Figure~\ref{fig:fig6} top).
However, testing the polar-cap scenario with these two pulsars is challenging
since their $B_S B_{\rm LC}$ values are almost the same (Figure~\ref{fig:fig6} bottom).

The NuSTAR data of J1849 enabled us to measure $E_{\rm SR}=58^{+43}_{-17}$\,keV for its SED,
thereby confirming the $E_{\rm SR}$-$B_S B_{\rm LC}$ correlation (Figure~\ref{fig:fig6} bottom).
The inclusion of the J1849 measurement in the $E_{\rm SR}$-$B_{\rm LC}^{7/2}$ trend
argues against the outer-gap scenario.
Measuring $E_{\rm SR}$ for more MeV pulsars would greatly contribute to a more detailed understanding of the pair-production mechanism.

\subsection{X-ray spectra of G32.6}
\label{sec:sec4_2}
G32.6 displays a spectrally hard X-ray emission ($\Gamma\approx 1.5$) in the inner region
($R<20''$ from the pulsar). We note that \citet{kh15} reported smaller values of $\Gamma$ in inner regions.
We speculate that the discrepancy in the $\Gamma$ measurements, albeit within the uncertainties, is caused by the contamination from the pulsar emission in the analysis of \citet{kh15}. The $\Gamma\approx 1.5$ spectrum indicates that the energy index of the electron distribution
is $p_1 \approx 2$, although the large uncertainty of $\Delta \Gamma\approx 0.5$ in this region poses challenges in accurately constraining $p_1$. Spatially-integrated SEDs can also offer an alternative method for determining $p_1$ of the uncooled electron distribution, as
it is reflected in the synchrotron SED in the IR-to-optical band or in the IC SED at energies $\lapp$10\,TeV. However, measurements in these bands are either lacking or of poor quality, hampering precise estimations of $p_1$. Moreover, the observed SED is a superposition of multi-zone emission components from various radial regions projected onto the observer's tangent plane. Thus, the intrinsic spectrum in the inner region might have a lower value of $p_1$. Nevertheless, if the inferred value of $p_1 \approx 2$ is indeed accurate, it may suggest that magnetic reconnection or shock drift acceleration plays a role in particle acceleration at the TS of PWNe \citep[e.g.,][]{Summerlin2012,Sironi2017}.

Alternatively, the $\Gamma\approx 1.5$ spectrum in the innermost region could be attributed to putative substructures such as the jet and counterjet (Figure~\ref{fig:fig2} (b)), as these features can exhibit spectrally hard emission \citep[e.g.,][]{Safi-Harb2008,AnMSH2014}. These substructures might potentially contain a distinct electron population. A deeper Chandra observation will facilitate more definitive detections 
and spectral characterization of these substructures, elucidating the origin of their hard spectra. This will, in turn, provide valuable clues about the mechanisms responsible for particle acceleration at TS of middle-aged PWNe.

At large distances from the pulsar (e.g., $R>20''$), the X-ray spectrum follows a $\Gamma\approx 2$ PL, similar to other PWNe. 
In comparison to the $\Gamma\approx 1.5$ spectrum in the inner region, this
indicates rapid particle cooling. 
As electrons travel outward within the inner region ($R\approx 20''$), they progressively lose kinetic energies. As a result, the emission peak of the highest-energy electrons will appear in the X-ray band and enable determining the flow speed and $B$ in the PWN. However, the spectral variation over the PWN was not measured with high significance with the current X-ray data, and further confirmation of this spectral softening is necessary with future observations. 

\subsection{SED modeling}
\label{sec:sec4_3}
We applied a multizone PWN model to infer the properties of the PWN G32.6 (Table~\ref{ta:ta3}).
While our model broadly reproduces the measurements (Figure~\ref{fig:fig5}),
the rapid change in $\Gamma$ in the inner regions, if real, is not very well explained.
At face value, the $\Delta \Gamma\approx 0.5$ can be explained by synchrotron cooling, which causes a change of the spectral slope of 1 (in ideal cases) between the distributions of the injected and cooled electrons. However, the `observed' spectrum is a superposition of emissions from many different radial regions, and thus the `observed' difference in $\Gamma$ would be diluted if observed by the current X-ray telescopes, and appear smaller than the emitted one (the inner vs. outer radial zones). This implies that the difference in the slopes of the electron distributions (relevant to the X-ray band) between $R\lapp 20''$ (injected) and $R\gapp 20''$ (cooled) needs to be greater than 1. This is difficult to reproduce within the model unless we employ a broken power-law distribution for the injected electrons \citep[such broken power laws may be required to explain large amounts of spectral steepening seen in several PWNe;][]{Chevalier2005}.

The parameter values in Table~\ref{ta:ta3} are not well constrained nor uniquely
determined, as the parameters are interdependent \citep[e.g., see][]{Park2023a}
and the quality of the measurements is rather poor; e.g., various values of $\eta_e$ can reproduce
the observed results, as mentioned in Section~\ref{sec:sec3_2}.
However, if the energy densities of the seed photons estimated with the {\tt GALPROP} model, despite its low angular resolution, are accurate, $B_0$ can be determined relatively well independently of the other parameters,
based on the shapes and amplitudes of the synchrotron and IC SEDs (Section~\ref{sec:sec3_2}). This, in turn, helps estimate $\gamma_{e,\rm max}$ in combination with the NuSTAR spectrum. Moreover, the estimation of $B_0$ along with our measurements of the brightness and $\Gamma$ profiles can help pin down $\alpha_V$ and $V_0$ if the PWN expansion rate is measured.

By fitting the radial profiles of the X-ray brightness and $\Gamma$,
we inferred the index of the bulk-flow speed to be $\alpha_V=-0.7$.
With this speed profile, electrons can be transported to 8.5\,pc, which is sufficient to explain
the observed X-ray emission zone with a size of 5\,pc. The softening of the X-ray spectrum with increasing distance from the pulsar requires diffusion, as pure advection models predict a constant $\Gamma$ profile in the inner regions and a rapid increase in the outer regions \citep[][]{Reynolds2003}. In our model, we used the diffusion coefficient of $D_0=10^{26}\rm \ cm^2\ s^{-1}$. In this case, electrons responsible for the TeV emission (e.g., off of IR seeds), with $\gamma_e\approx 10^7$, can move out to a distance of $\sim$10\,pc through diffusion and advection, which
can explain the TeV emission size of $0.09^\circ$. In this region located outside the X-ray PWN, the electrons will cool primarily through IC emission and propagate further outward by diffusion. 
Assuming a spatially uniform ISRF with $u_{\rm IR}=0.5\rm \ eV\ cm^{-3}$ and the ISM $B$ ($B_{\rm ISM}$) of 1$\mu$G, the cooling timescale of electrons with $\gamma_e\approx 10^7$ is estimated to be 12\,kyr. Then, our model predicts that the TeV-emitting electrons will reach a distance of $\sim 30$\,pc from the pulsar in 12\,kyr and form an extensive TeV halo based on the assumed diffusion law (Equation~(\ref{eq:diffusion})). However, this estimate is subject to substantial uncertainty since the values of $u_{\rm IR}$ and/or $B_{\rm ISM}$ may differ significantly.

Our estimate of $\gamma_{e,\rm max}$ corresponds to a maximum electron energy
of $E_{e,\rm max}\approx$400\,TeV which is comparable to 740\,TeV inferred from modeling
the TeV emission of G32.6 \citep[][]{Amenomori2023}.
With the lack of measurements at $>$20\,keV, our estimation of $E_{e,\rm max}$ should be regarded as a lower limit.
Additionally, if the $\Gamma\approx 1.5$ spectrum in the innermost region represents the intrinsic spectrum of the PWN rather than putative substructures, our model predicts a spectral curvature at $\sim$5\,keV.

\section{Summary}
\label{sec:sec5}
Here is a summary of our work.
\begin{itemize}
\item We measured the on$-$off spectrum of J1849 and found that its spectrum exhibits
curvature with a peak at 60\,keV, which supports the scenario that energetic electron-positron pairs in pulsar magnetosphere
are generated by polar-cap cascades \citep[][]{Harding2017}.
\item The off-pulse emission of J1849 has a very similar spectral shape to the on$-$off one and
contains $\sim$20\% flux of the `on$-$off' pulse emission. This characterization allowed us to measure the broadband X-ray spectrum of the PWN.
\item Our multizone emission modeling found that G32.6 can accelerate electrons to $\gapp$400\,TeV. Additionally, the model predicts that the TeV emitting electrons may propagate out to a distance of $\sim$30\,pc from the pulsar in $\sim$10\,kyr, although these values may substantially alter depending on the properties ($B_{\rm ISM}$ and $u_{\rm IR}$) of the ISM.
\end{itemize}

While we were able to accurately measure the emission properties of J1849, characterizing the PWN emission was challenging due to its faintness and strong contamination from the pulsar. Consequently, inferring the PWN properties using the model was difficult. More sensitive X-ray data, to be collected by future X-ray observatories with high angular resolution and large collecting area \citep[e.g., AXIS and HEX-P;][]{Mushotzky2019,Madsen2019}, would be highly beneficial. Additionally, further theoretical studies that encompass not only the spatial variations but also the temporal variations of PWN properties \citep[such as flow speed, $B$, etc;][]{Gelfand2009,Bandiera2023} will be crucial for understanding high-energy electrons within the Galaxy.

\bigskip
\bigskip

\begin{acknowledgments}
This work used data from the NuSTAR mission, a project led by the California Institute of Technology,
managed by the Jet Propulsion Laboratory, and funded by NASA. We made use of the NuSTAR Data
Analysis Software (NuSTARDAS) jointly developed by the ASI Science Data Center (ASDC, Italy)
and the California Institute of Technology (USA).
JP acknowledges support from Basic Science Research Program through the National
Research Foundation of Korea (NRF) funded by the Ministry of Education (RS-2023-00274559).
AB acknowledges support from Japan Society for the Promotion of Science Grants-in-Aid
for Scientific Research (KAKENHI) Grant Numbers  JP23H01211.
Support for this work was partially provided by NASA through NuSTAR Cycle 6
Guest Observer Program grant NNH19ZDA001N.
This research was supported by the National Research Foundation of Korea (NRF)
grant funded by the Korean Government (MSIT) (NRF-2023R1A2C1002718).
We thank the referee for comments that helped improve the clarity of the paper.
\end{acknowledgments}

\bigskip

\vspace{5mm}
\facilities{CXO, XMM-Newton, NICER, Swift, NuSTAR}
\software{HEAsoft \citep[v6.31;][]{heasarc2014}, CIAO \citep[v4.14;][]{fmab+06},
XMM-SAS \citep[20211130\_0941;][]{xmmsas17}, XSPEC \citep[v12.12;][]{a96}}
%FermiPy \citep[v1.0.1;][]{Wood2017}

%\vspace{5mm}
%\facilities{HST(STIS), Swift(XRT and UVOT), AAVSO, CTIO:1.3m, CTIO:1.5m,CXO}
%\software{astropy \citep{2013A&A...558A..33A},  
%          Cloudy \citep{2013RMxAA..49..137F}, 
%          SExtractor \citep{1996A&AS..117..393B}
%          }

\bibliographystyle{apj}
\bibliography{GBINARY,BLLacs,PSRBINARY,PWN,STATISTICS,FERMIBASE,COMPUTING,INSTRUMENT,ABSORB,PULSARS,MAGNETAR,CATALOG}
%\bibliography{J1849rev.bbl}
\expandafter\ifx\csname natexlab\endcsname\relax\def\natexlab#1{#1}\fi

\end{document}